\documentclass[preprint,12pt, a4paper]{elsarticle}

\usepackage{amsmath,amsbsy}

\newcommand*\Bell{\ensuremath{\boldsymbol\ell}}

\usepackage{subfig}

\biboptions{round, authoryear}

\usepackage[nomarkers, nolists,figuresonly]{endfloat}

\journal{ICARUS}

\begin{document}

	\begin{frontmatter}

		\title{3D Meteoroid Trajectories}
		
		\author{Eleanor K. Sansom\corref{cor1} }
		\author{Trent Jansen-Sturgeon\corref{cor1} }
		\author{Mark G. Rutten\corref{cor2} }
		\author{Hadrien A. R. Devillepoix\corref{cor1}}
		\author{Phil A. Bland\corref{cor1}}
        \author{Robert M. Howie\corref{cor3}}
        \author{Morgan A. Cox\corref{cor1}}
		\author{Martin C. Towner\corref{cor1}}
        \author{Martin Cup\'ak\corref{cor1}}
        \author{Benjamin A. D. Hartig\corref{cor3}}

		\cortext[cor1]{School of Earth and Planetary Sciences, Curtin University, GPO Box U1987, Perth, WA 6845, Australia}
		\cortext[cor2]{Defence Science and Technology Group, Edinburgh, SA 5111, Australia}
		\cortext[cor3]{Department of Mechanical Engineering, Curtin University, GPO Box U1987, Bentley, Perth, WA 6845, Australia}
		
		
		\begin{abstract}
			Meteoroid modelling of fireball data typically uses a one dimensional model along a straight line triangulated trajectory. The assumption of a straight line trajectory has been considered an acceptable simplification for fireballs, but it has not been rigorously tested. The unique capability of the Desert Fireball Network (DFN) to triangulate discrete observation times gives the opportunity to investigate the deviation of a meteoroid's position to different model fits. 
			Here we assess the viability of a straight line assumption for fireball data in two meteorite-dropping test cases observed by the Desert Fireball Network (DFN) in Australia -- one over 21 seconds (\textit{DN151212\_03}), one under 5 seconds (\textit{DN160410\_03}). We show that a straight line is not valid for these two meteorite dropping events and propose a three dimensional particle filter to model meteoroid positions without any straight line constraints. The single body equations in three dimensions, along with the luminosity equation, are applied to the particle filter methodology described by \citet{Sansom2017}. 
            Modelling fireball camera network data in three dimensions has not previously been attempted. 
            This allows the raw astrometric, line-of-sight observations to be incorporated directly. 
        	In analysing these two DFN events, the triangulated positions based on a straight line assumption result in the modelled meteoroid positions diverging up to $3.09\, km$ from the calculated observed point (for \textit{DN151212\_03}). Even for the more typical fireball event, \textit{DN160410\_03}, we see a divergence of up to $360$\,m.
        	As DFN observations are typically precise to $<100$\,m, it is apparent that the assumption of a straight line is an oversimplification that will affect orbit calculations and meteorite search regions for a significant fraction of events. 
            	
		\end{abstract}
		
			
		
	\end{frontmatter}

\section{Introduction} 
When meteoroids pass through the Earth's atmosphere the luminous phenomena produced can be characterised by its brightness, increasing from meteor to fireball to bolide \citep{Ceplecha1998}. 
Meteors are typically associated with cometary dust and burn up high in the atmosphere. 
Fireballs tend to be slower than meteors and more likely of asteroidal origin. These lower entry velocities allow meteoroids to penetrate deeper into the atmosphere, with longer trajectories likely to be influenced by its increasing density. 
Fireballs are particularly significant as they are frequent enough for dedicated camera networks to capture regularly, whilst still having the potential for objects to survive entry and drop meteorites to Earth. 
Modelling of fireball trajectories for orbit analysis and meteorite recovery is typically based on a straight line assumption \citep{McCrosky1965, Spurny2012, Brown1994, Hildebrand2006}. 
The synchronised astrometric observations acquired by the Desert Fireball Network (DFN; \citealt{howie2017deb}) provide a unique opportunity to test this assumption. 
This work analyses two fireball test cases and introduces a modelling technique that uses raw observational data to estimate a trajectory without the need for pre-triangulated data. 
Although DFN data are used, they are simply to illustrate the issues surrounding the straight line assumption and the functionality of the 3D particle filter technique presented.

\subsection{Modelling and observing fireball trajectories}
Determining the potential of a fireball to produce a meteorite involves a trajectory analysis of each individual event. 
The meteoroid can be modelled based on the single body theory of meteoroid dynamics -- 
a set of continuous differential equations 
representing the evolution of a meteoroid's behaviour as it passes through the atmosphere \citep{Hoppe1937, Baldwin1971, Sansom2017}.
This is, however, a simplified theory and does not explicitly include any disruptions to the  body. 
Furthermore, many of the trajectory parameters are unknown and assumptions must be made, or models used, to determine their values. 

Models such as those used by \citet{Ceplecha2005} and \citet{Kikwaya2011} 
apply a least squares methodology to determine the characteristics of a meteoroid during its flight based on positional observations and light curves. 
A least squares approach however does not rigorously examine the uncertainties in observations, or the limitations posed by the single body model applied, when evaluating errors. Typically the observational residuals to a straight line fit are quoted as positional uncertainties for the trajectory. This is not valid as the errors induced by using any model must be incorporated.

Even though meteor ablation models \citep{Campbell-Brown2004, Kikwaya2011}  expand on the single body equations for ablation 
by including thermal fragmentation mechanisms, their application is limited to small meteor-producing bodies ($10^{-12} \text{ to } 4\times10^{-5}\,kg$ / $10\,\mu m \text{ to } 2\,mm$ ;  \citealt{Campbell-Brown2004}).

Hydrodynamic numerical models (such as SOVA \citep{Shuvalov1999} and the model of \citealt{Shuvalov2002}) focus on external processes for modelling the interaction and propagation of shock waves through the atmosphere caused by hypersonic flight of bolides \citep{Artemieva2016}. 
These models do not use raw observational data and are 
computationally expensive procedures \citep{Artemieva2009}.
For this reason a pragmatic approach, such as the particle filter technique used by \citet{Sansom2017} (after \citealt{Ristic2004}), is favoured to characterise meteoroid atmospheric entry of large fireball network data sets. 
The Monte Carlo technique of \citet{Sansom2017} iteratively estimates the state of the trajectory system at each observation time. It does not aim to fit the entire trajectory at once. This removes the assumptions and limitations of normal fitting techniques that may force the simplified single body equations to model this more complex system.
Despite the particle filter using these equations as a base model, the adaptive approach uses the observations and appropriate covariances to incorporate, to some extent, unmodelled processes (such as fragmentation). The nature of this technique allows a broad range of trajectory parameters (including densities, shapes and ablation parameters) to be explored, and favourable values to be identified, in a more robust way than a brute force least squares approach. 

Beside modelling a meteoroid's dynamic trajectory, it is possible to relate the mass loss of the body along the trajectory to the observed brightness of the event, as a portion of the kinetic energy loss is transformed into visible light \citep{Ceplecha1996}. This can be modelled following the differential equation 
\begin{equation}\label{eqn:I}
    I = -\tau\left(1+\frac{2}{\sigma v^2}\right)\frac{v^2}{2}\frac{dm}{dt}\times 10^{7}.
\end{equation}
The luminosity, $I$, is typically referred to in $erg\,s^{-1}$ but is given here in SI units of Watts (and thus introducing the conversion factor of $10^7\, W\,s\,erg^{-1}$). The percentage of energy that is converted to radiation is quantified by the luminous efficiency, $\tau$. 
$v$ and $m$ are the velocity and mass of the meteoroid with $t$ being the observation time and $\sigma$ the ablation parameter.

As fireball observations by the DFN are only in the visible wavelengths, as is typical for such networks, the luminosity values need to be adjusted depending on the meteoroid temperature. A value of $1.5 \times 10^{10}$ is used to relate a typical source temperature of $4500 \,K$ to the luminosity in the visual pass-band, $I_v$ \citep{Ceplecha1996}.
If the observed brightness values can be expressed in absolute visual stellar magnitudes,$M_v$, then a comparison may be made to models using Equation \eqref{eqn:I} by:
\begin{equation}\label{eqn:Iv}
    \cfrac{I}{1.5 \times 10^{10} } = I_v = 10^{-0.4 M_v }.
\end{equation}
Incorporating the fireball's calculated luminosity into the particle filter methodology is able to provide an additional observation to the filter, helping to further constrain mass loss estimates. 
The luminosity of a fireball can be calculated based on the long exposure images or by calibrating the measurements of an external device such as a radiometer.

When a fireball is captured by multiple Desert Fireball Network remote observatories, each camera image is calibrated using the background star field to determine an astrometric azimuth and elevation for positions along the fireball trail. 
This method of calibration (detailed by \citet{2018arXiv180302557D}) accounts well for any effects of atmospheric refraction, and the uncertainty introduced by the calibration is typically less than 1 arcminute.
The DFN camera systems encode absolute timing in fireball trajectories using a modulated liquid crystal shutter within the lens of each camera \citep{howie2017deb}. The De-Brujn encoding embedded within the fireball trail itself is synchronised across the network via GNSS. This gives us the unique capability of individually triangulating meteoroid positions for every discrete time-step with multi-station observations. This has not hereto been possible. 
Despite the uncertainties, with correct error analysis this triangulation of discrete observation times can give us `ground truth' positions of the meteoroid with which we can compare different approaches to meteoroid trajectory analyses.

Here we assess the viability of a straight line assumption for fireball data by comparing straight line positions to those calculated using this unique triangulation capability of the DFN.
We also propose a three dimensional particle filter to model meteoroid positions without any straight line constraints.
The single body equations in three dimensions, along with the luminosity equation, are applied to the particle filter methodology described by \citet{Sansom2017}. 
In doing this, the observations used by the filter to update the state vector are permitted to be in the form of the raw line-of-sight observations in azimuth and elevation as well as luminosities (where available). This drops the simplifying assumption of a straight line trajectory entirely, as particles are free to move in three dimensional space. Error propagation is thorough as the filter considers the observational uncertainties in each azimuth and elevation individually as well as considers trajectory model limitations. 

The better the understanding we have of the final state of a meteoroid, and the uncertainties throughout the modelling phase, the more confidence we have in predicted fall regions. This may significantly influence decisions regarding the feasibility of ground-based searches for meteorites.

\section{Assessing the limitations of the straight line assumption}\label{sec:slls}

Historically, there have been two predominant meteoroid triangulation methods; the method of planes \citep{Ceplecha1987} and the straight line least squares (SLLS) method \citep{Borovicka1990}. 
The method of planes involves finding the best fit, 2D plane for each observatory that contains both the observatory location and the line-of-sight meteoroid observations. The intersection of multiple planes defines the trajectory; in the case of more than two observatories, this will result in multiple trajectory results which are then averaged in practice.
The straight line least squares method on the other hand determines a best fit, straight line radiant for the trajectory considering all the raw observations at once. This is done by minimising the angular difference between the observed lines of sight and the line joining the observatory to the closest corresponding point along the best fit radiant line. 
By assuming a straight line trajectory, this effectively destroys any subtleties in the data by forcing it to fit what may potentially be an oversimplified model. 
The straight line assumption may be an acceptable simplification for some events, especially short, fast meteors, but may not always be valid for longer fireballs with significant deceleration and should be tested.

A least squares approach however does not rigorously examine the uncertainties in observations, or the limitations posed by the single body model used, when evaluating errors. Typically the observational residuals to a straight line fit are quoted as positional uncertainties for the trajectory. This is not valid as the errors induced by using any model must be incorporated.
Despite the decrease in residuals when considering the upper sections of the trajectory only (observations of the fireball above 50 km), it must be noted that this is not a good measure of the true trajectory uncertainties as the model errors are not taken into account.

The reference frame in which the straight line is fitted also needs be considered for such long, decelerating events and is rarely discussed. It is expected that a fireball trajectory is approximately straight in an inertial reference frame only, and that Earth rotation effects will cause an apparently curved path for an observer on the ground. This requires accurate timing throughout the meteoroid flight. 
Although \citet{Ceplecha1987} adjust entry vectors for both Earth rotation and gravity, this is intended to correct the heliocentric orbit beyond the sphere of influence of the Earth.
The SLLS method of \citet{Borovicka1990} allows the incorporation of time differences between measurements to account for Earth rotation effects, though it is not a requirement of the method; the authors even state that the local sidereal time of the observer is usually assumed to be constant throughout a meteor's flight.
For short events that do not show any significant deceleration, it is unlikely that these effects would be noticeable within the error of the observations. For fireballs that are longer and show significant deceleration however, this may no longer hold true. 
Most trajectory analyses of recent fireball events \citep{Brown2011, Borovicka2013kosice, Borovicka2015Krizevci, Spurny2017Atmospheric2016} cite the SLLS of \citet{Borovicka1990} as the method of trajectory determination, though it is not made apparent in every case which considerations have been made. Uncertainties in triangulated positions are also often quoted as the residuals to the straight line \citep{Spurny2010jen, Borovicka2013kosice, Borovicka2015Krizevci, Spurny2017Atmospheric2016} fit without taking into consideration the error of the straight line model and are therefore not a true representation of the trajectory uncertainty.

\subsection{Point-wise triangulation}\label{sec:pw}

\begin{figure}
    \centering
    \includegraphics[width=1\linewidth]{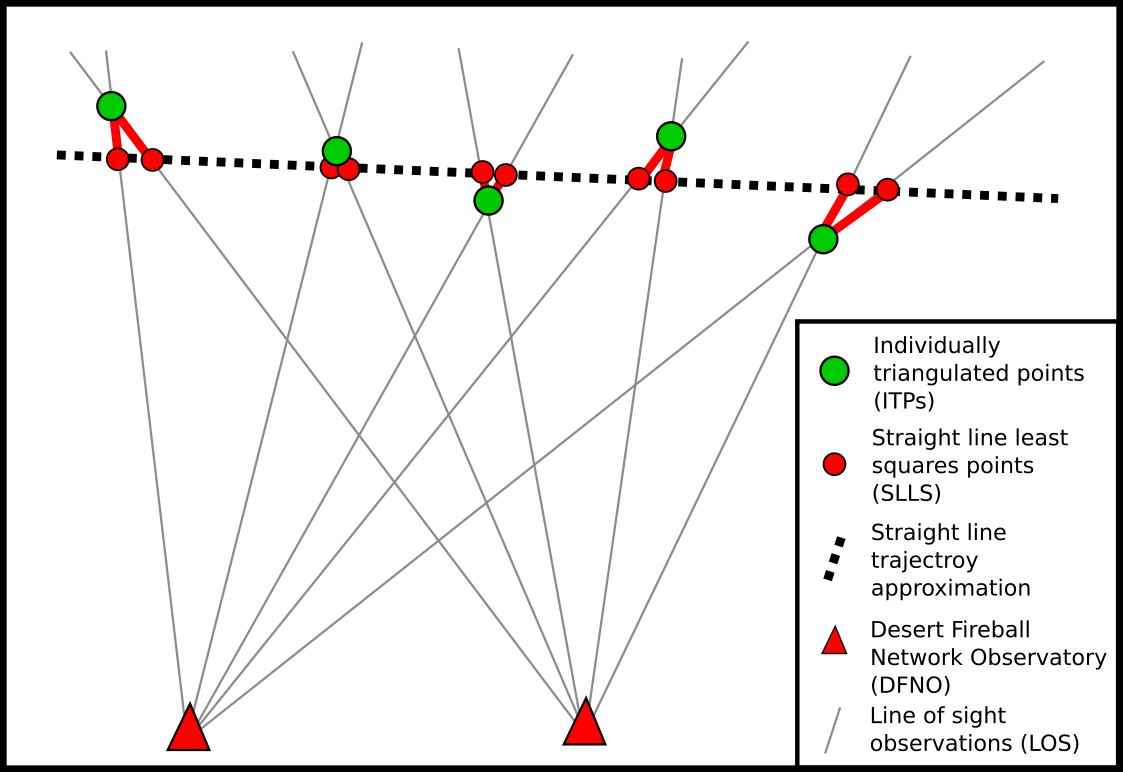}
    \caption{Schematic representation (not to scale) of five unique time steps observed by two DFNOs. Difference between the straight line least squares trajectory points (red) and the individually triangulated positions (green) are highlighted in red.}
    \label{fig:cam_schem}
\end{figure}

The unique method used by the DFN camera systems to encode absolute timing in fireball trajectories is synchronised across the network via GNSS. The instantaneous meteoroid position for a given time step can therefore be evaluated using what we here refer to as a \textit{point-wise triangulation} (schematically illustrated in Figure \ref{fig:cam_schem}). 
Point-wise triangulation estimates the meteoroid position, $\Bell$, by minimising the angular separation, $\theta$, between the \textit{calculated} line-of-sight unit vector to $\Bell$ and the \textit{observed} line-of-sight unit vector, $\mathbf{z}^{n}$ for each observatory, $\mathbf{O}^{n}$  (where $\mathbf{z}^{n}$, $\Bell$ and $\mathbf{O}^{n}$ are in an ECEF rectangular geocentric coordinate system).

\begin{equation}\label{eqn:theta}
\theta =  \sqrt{ \sum_n \left[ \arccos\left(\cfrac{\Bell-\mathbf{O}^n}{||\Bell-\mathbf{O}^n||} \,\,\bullet\, \mathbf{z}^{n} \right) \right]^{2} }
\end{equation}

The resulting individually triangulated positions (ITPs) are used as a reference for comparison of trajectory models. 

\subsection{Introducing two fireball test cases}
Here we detail two fireballs observed by the Desert Fireball Network and assess the appropriateness of a straight line trajectory fit for these cases. 

\subsubsection{Case 1: DN151212\_03 -- long, shallow}
\begin{figure}
	\centering
	\includegraphics[width=1\linewidth]{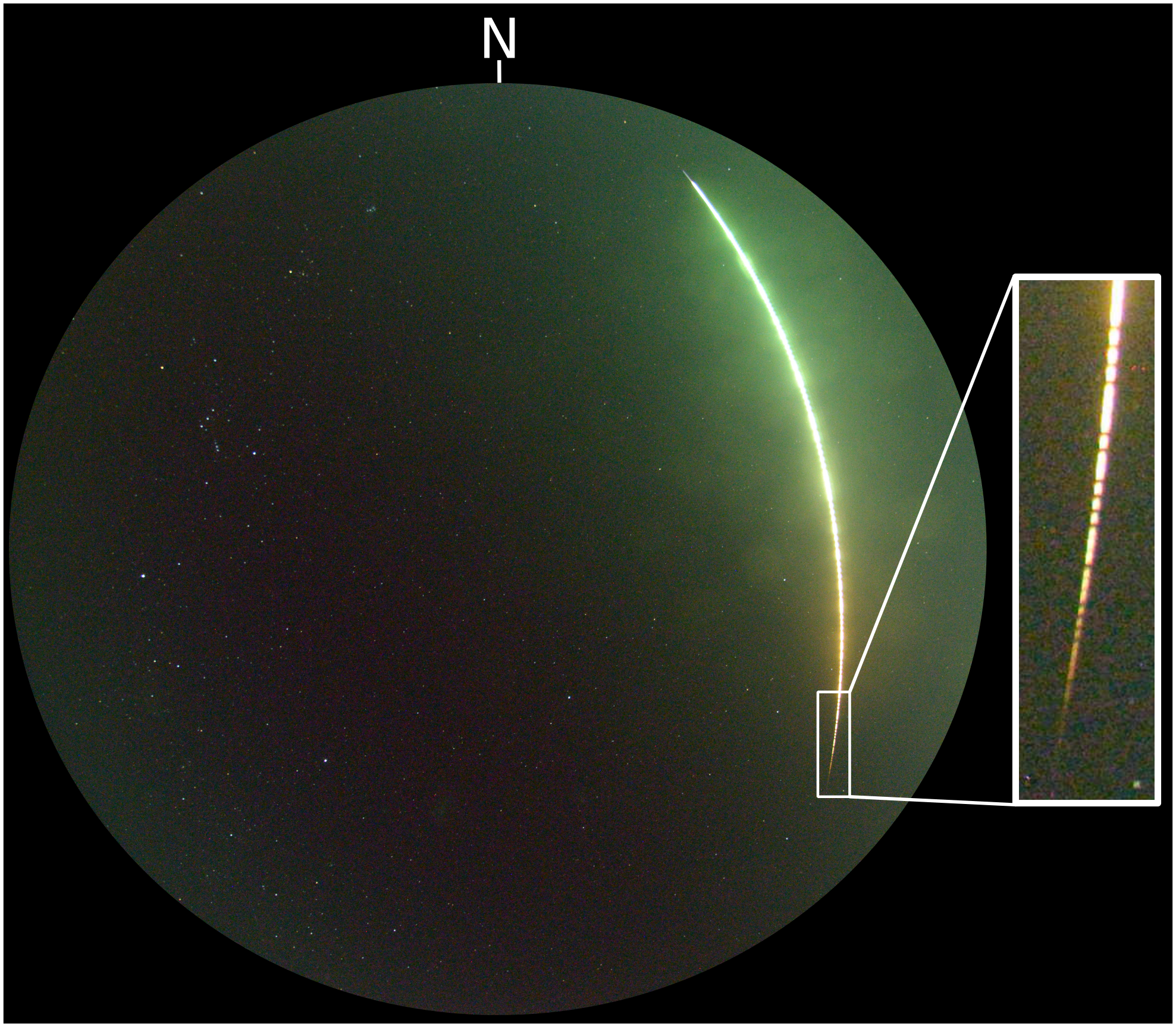}
	\caption{\textit{DN151212\_03} fireball as seen from Etadunna Station, South Australia, travelling from North (top) to South (bottom) with a final recorded point at 11:36:45.526 UTC. Calibration with background stars determines azimuth and elevation of trajectory points.}
	\label{fig:151212_39}
\end{figure}

On the 12th of December 2015, at 11:36:23.826 UTC, a
$>21$ second long fireball over South Australia was captured by five DFN observatories east of Kati Thanda (hereafter referred to as event \textit{DN151212\_03}). 
DFN systems at this time captured two 25 second exposures with timing every minute, and the fireball was split over two consecutive images. The fireball appeared in the last $\sim$2 seconds of the first exposure, was unobserved during the gap between exposures, and further captured for another $\sim$14 seconds in the second exposure, with a final observation time at 11:36:45.526 UTC. Figure \ref{fig:151212_39} shows the second exposure captured by the observatory closest to the terminal point (DFNO\_39). The modulation of the liquid crystal shutter used to encode absolute and relative timing can be seen as long and short dashes along the trajectory. 

\begin{figure}
    \centering
    \includegraphics[width=\textwidth]{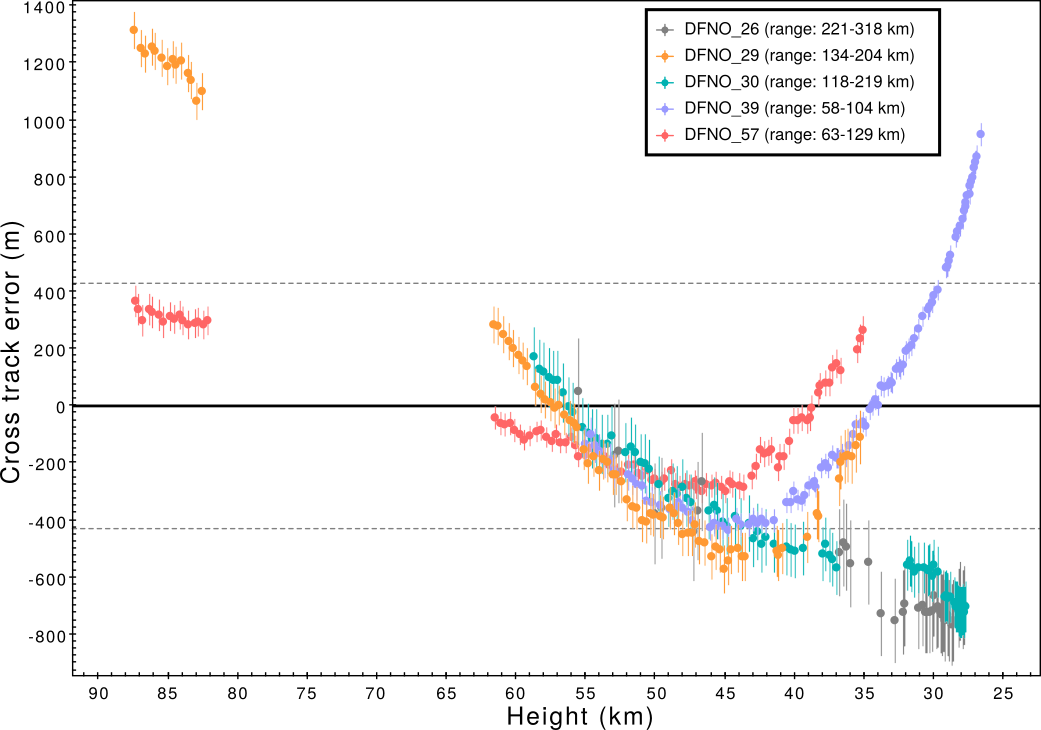}
    \caption{Cross track residuals for individual station observations of \textit{DN151212\_03} to a straight line trajectory fit in an inertial (ECI) reference frame. Range values in legend are the minimum and maximum distances of a station to the fireball trajectory. Error bars on observations are $1\sigma$ astrometric errors propagated by the corresponding range. The gap between 82-62 km corresponds to the $\sim5$ second gap between exposures. }
    \label{fig:bv_xtrack}
\end{figure}

Initially the entire 21.7 second trajectory was fitted using the straight line least squares (SLLS) method following \citet{Borovicka1990}. As absolute timing is known to a high accuracy, this is preferably performed in an Earth centred, inertial (ECI) reference frame, though a non-inertial (Earth centred, Earth fixed; ECEF) solution is also calculated to assess the variation in fits.  
A 1D extended Kalman smoother trajectory analysis \citep{Sansom2015} on these straight line data estimates the trajectory parameters. The results for both reference frames are given in Table \ref{tab:bv_slls}. 
\begin{table}[]
    \centering
    \begin{tabular}{l|c|c}
         \textit{DN151212\_03} (full)& ECI & ECEF \\ \hline\hline
         entry radiant -- RA ($\,^\circ$) &$23.77\pm0.37$&$26.43\pm0.45$ \\\hline
          entry radiant -- DEC  ($\,^\circ$) &$46.17\pm0.13$&$46.12\pm0.09$ \\\hline
         initial height (km) & $87.7\pm0.1$
                        & $88.5\pm0.1$\\\hline
         initial velocity (km s$^{-1}$) & $13.09\pm0.07$ 
                          & $13.15\pm0.07$\\ \hline
         initial mass (kg)& $35\pm2$ 
                          & $33\pm2$\\ \hline 
         entry slope, $\gamma_e$ ($^\circ$) & $16.4$
                     & $16.5$\\ \hline
         final height (km) & $26.5\pm0.1$
                      & $26.4\pm0.2$\\ \hline
         final velocity (km s$^{-1}$) & $2.63\pm0.44$
                        &$2.4\pm0.5$\\\hline
        final mass (kg)& $2.0\pm0.2$
                      & $1.9\pm0.3$\\ \hline
    \end{tabular}
    \caption{Straight line least squares (SLLS) trajectory triangulated in either an inertial (ECI) or non-inertial (ECEF) reference frame for all observations of event \textit{DN151212\_03}. Trajectory characteristics (height, velocity, mass) are estimated using an extended Kalman smoother in one dimension on these straight line data. Entry slope is given as an angle from horizontal. The angular separation between the two radiants is $1.84^\circ$.}
    \label{tab:bv_slls}
\end{table}
The cross-track residuals of individual camera observations to the straight line fit (in ECI) can be seen in Figure \ref{fig:bv_xtrack}.
The 1.8$^\circ$ difference in entry radiant between inertial and non-inertial reference frames (Table\ref{tab:bv_slls}) shows the importance of including Earth rotation effects. 
The $>1\,km$ residuals show this is an inappropriate fit to the trajectory.

\begin{figure}
    \centering
    \includegraphics[width=\textwidth]{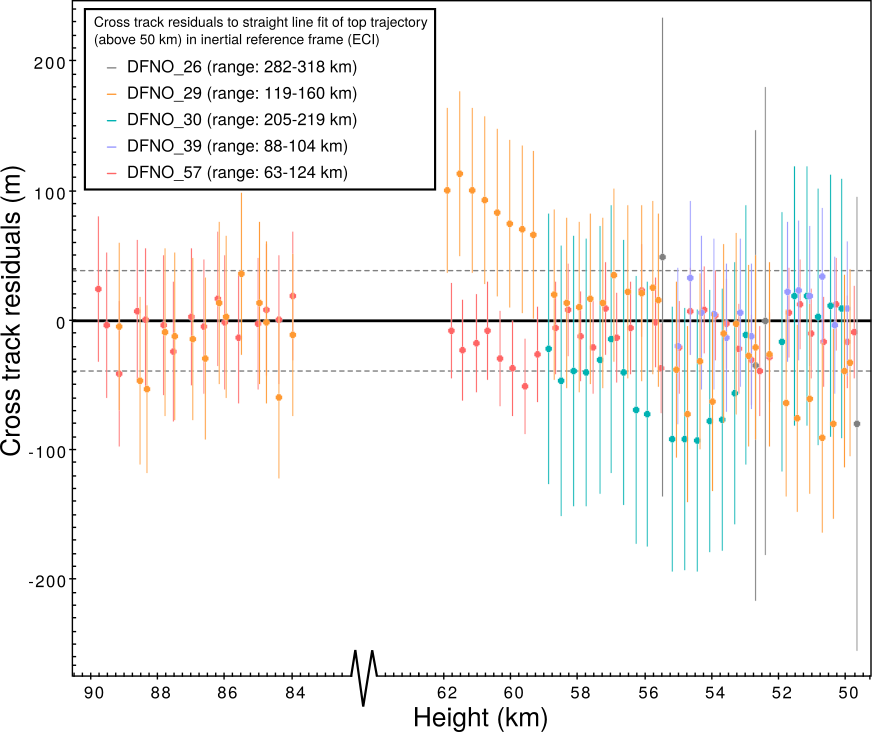}
    \caption{Cross track residuals for the upper half of the \textit{DN151212\_03} trajectory to a straight line fit in an inertial (ECI) reference frame. Only observations of the fireball above $50\,km$ were used. Error bars on observations are $1\sigma$ astrometric errors propagated by the corresponding range. }
    \label{fig:bv_xtrack_tops}
\end{figure}

\begin{figure}
    \centering
    \includegraphics[width=\textwidth]{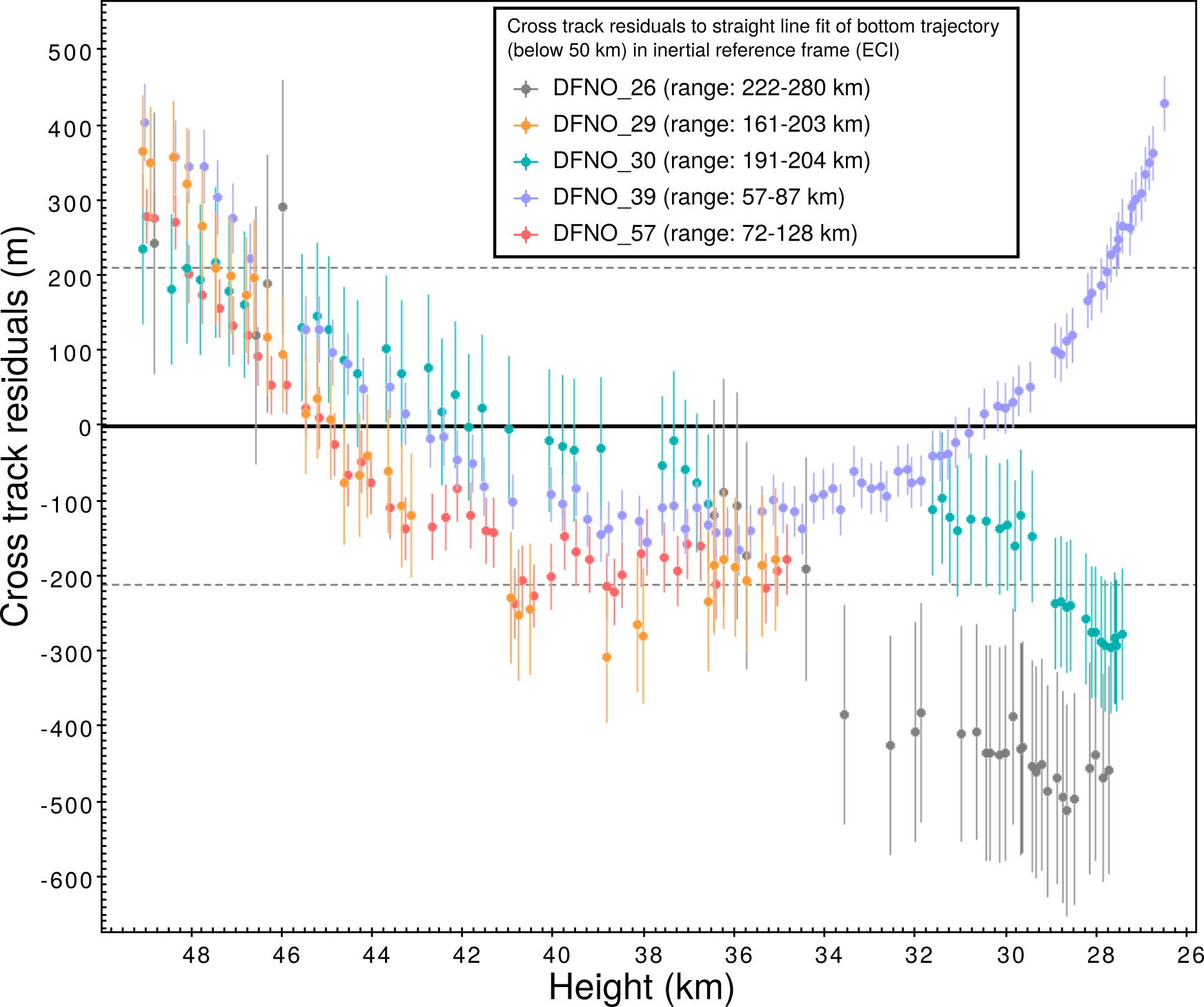}
    \caption{Cross track residuals for the lower half of the \textit{DN151212\_03} trajectory to a straight line fit in an inertial (ECI) reference frame. Only observations of the fireball below $50\,km$ were used. Error bars on observations are $1\sigma$ astrometric errors propagated by the corresponding range. }
    \label{fig:bv_xtrack_bots}
\end{figure}

\begin{figure}
    \centering
    \includegraphics[width=\textwidth]{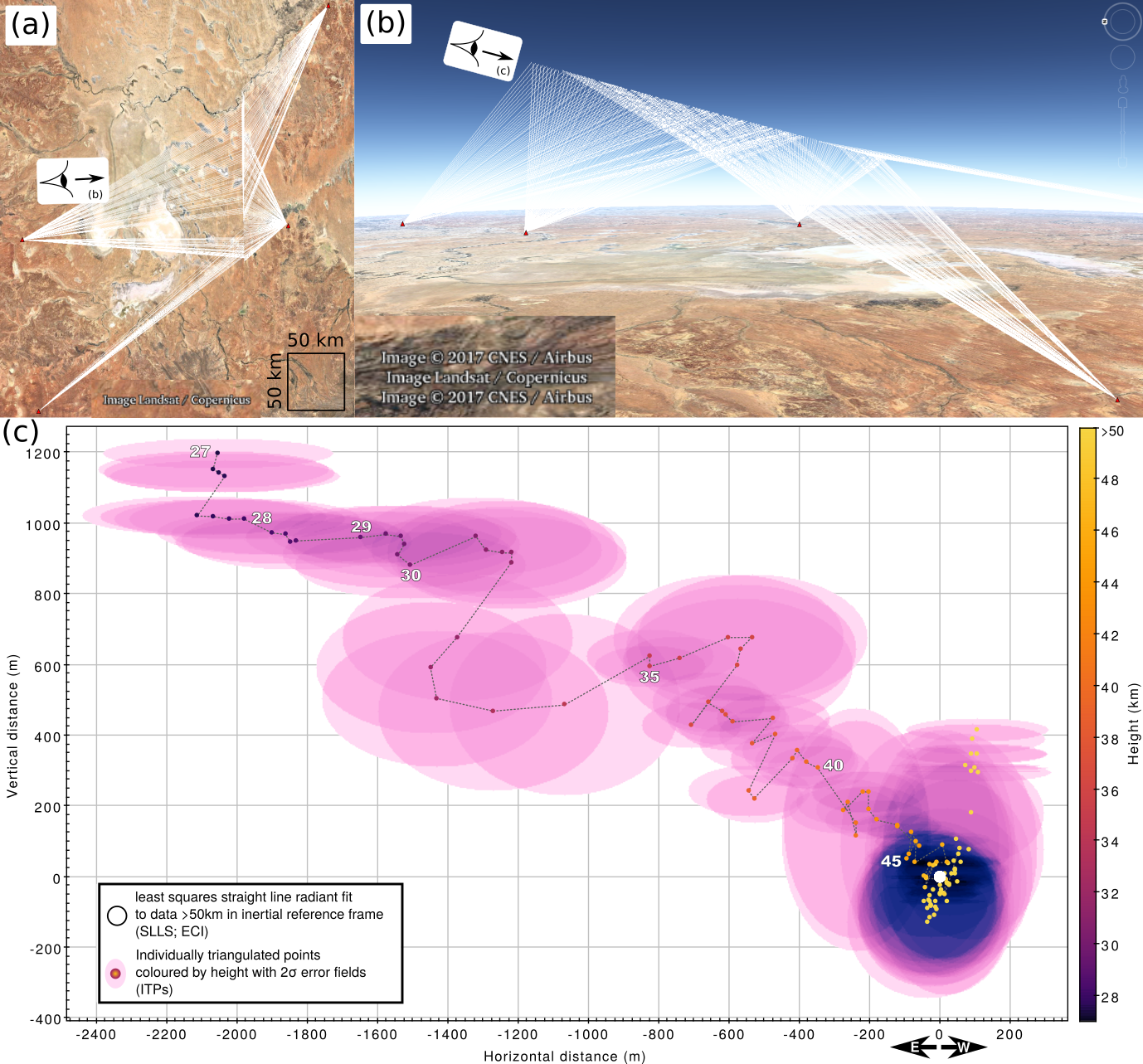}
    \caption{``Down-line" view (c) as seen by an observer looking down the ECI straight line radiant (illustrated by (a)--(b)) calculated using the top half of the trajectory (points $>50\,km$; see Figure \ref{fig:bv_xtrack_tops}). This results in the ECI trajectory stacking to a single point (white) whereas the individually triangulated positions (ITPs; coloured points) are projected onto the viewing plane. This plane is normal to the straight line trajectory with the x-axis aligned with the Earth horizontal, and inclined from true vertical by the cosine of the trajectory slope. 
    From this down-line view the ITPs help to illustrate the true non-linearity of the path taken by the meteoroid. (Google Earth image credit: Landsat/Copernicus/CNES/Airbus).}
    \label{fig:bv_downline}
\end{figure}

By considering only the observations above $50\,km$ we hope to improve the fit and calculate a more realistic entry radiant. The cross-track residuals to the straight line fit (in ECI) for its subset are seen in Figure \ref{fig:bv_xtrack_tops}.
The decreases in observation residuals to the straight line model shows a significantly improved fit, providing a more reliable entry radiant ($0.83^\circ$ difference between Table \ref{tab:bv_slls} and \ref{tab:bv_slls_tops} values). Despite the decrease in residuals, it must be noted that this is not a good measure of the true trajectory uncertainties as the model errors are not taken into account. 
Updated entry parameters given in Table \ref{tab:bv_slls_tops} are again calculated using an EKS \citep{Sansom2015}, which incorporates both observational and model errors in the quoted uncertainties. Non-inertial SLLS (ECEF) results are also quoted to highlight that despite improved fits in both reference frames, the radiant angles are still nearly 1.8$^\circ$ apart.
\begin{table}[]
    \centering
    \begin{tabular}{l|c|c}
         \textit{DN151212\_03} ($>50\,km$) & ECI &  ECEF \\ \hline\hline
         entry radiant -- RA ($\,^\circ$) &$24.18\pm0.03$&$26.69\pm0.06$ \\\hline
          entry radiant -- DEC  ($\,^\circ$) &$45.39\pm0.01$&$45.58\pm0.01$ \\\hline
          initial height (km) & $89.99\pm0.02$
                        & $89.97\pm0.02$\\\hline
         initial velocity (km s$^{-1}$) &  $13.52\pm0.06$
                          & $13.47\pm0.05$\\ \hline
         entry slope, $\gamma_e$ ($^\circ$) & 17.1  & 17.1\\ \hline
    \end{tabular}
    \caption{Straight line least squares (SLLS) trajectory triangulated in either an inertial (ECI) or non-inertial (ECEF) reference frame for observations of event \textit{DN151212\_03} above 50 km only. Trajectory characteristics (height, velocity, mass) are estimated using an extended Kalman smoother in one dimension on these straight line data. The angular separation between the two radiants is $1.77^\circ$.}
    \label{tab:bv_slls_tops}
\end{table}
A similar exercise can be performed with the lower half of the trajectory (observations $<50\,km$). Figure \ref{fig:bv_xtrack_bots} shows that a SLLS fit to these data still does not well represent the data and is little improved from Figure \ref{fig:bv_xtrack}. 
Rather than continuing to chop the trajectory into increasingly small segments, we can observe the path of the ITPs relative to the entry radiant calculated in Table \ref{tab:bv_slls_tops}. Figure \ref{fig:bv_downline} is a view looking down the ECI entry radiant (white point). This ``down-line" view projects all points onto the plane normal to the straight line trajectory, resulting in the ECI trajectory stacking to a single point. The x-axis is truly horizontal, and as the meteoroid travelled from North to South, negative deviations are to the East, while positive deviations are to the West. The y-axis is the deviation above and below the straight line trajectory and values can be translated to true vertical using the cosine of the trajectory slope. From this down-line view, we can gain an understanding of the true non-linearity of the \textit{DN151212\_03} meteoroid trajectory; the lower half is not randomly scattered around a straight line as points above 50 km are, rather they show a distinct lateral deviation to the East. This also shows that, despite the 21.7 second trajectory theoretically accumulating a  2.1\,km vertical drop due to gravity, this is not the cause of the deviation from a straight line.  

This fireball represents an interesting case, showing that effects other than Earth rotation and gravity are involved in significantly influencing trajectories. This long, shallow trajectory however is certainly not a regular event. Next we perform a similar analysis on a more typical fireball case.

\subsubsection{Case 2: DN160410\_03 -- typical event}
\begin{figure}
    \centering
    \includegraphics[width=\textwidth]{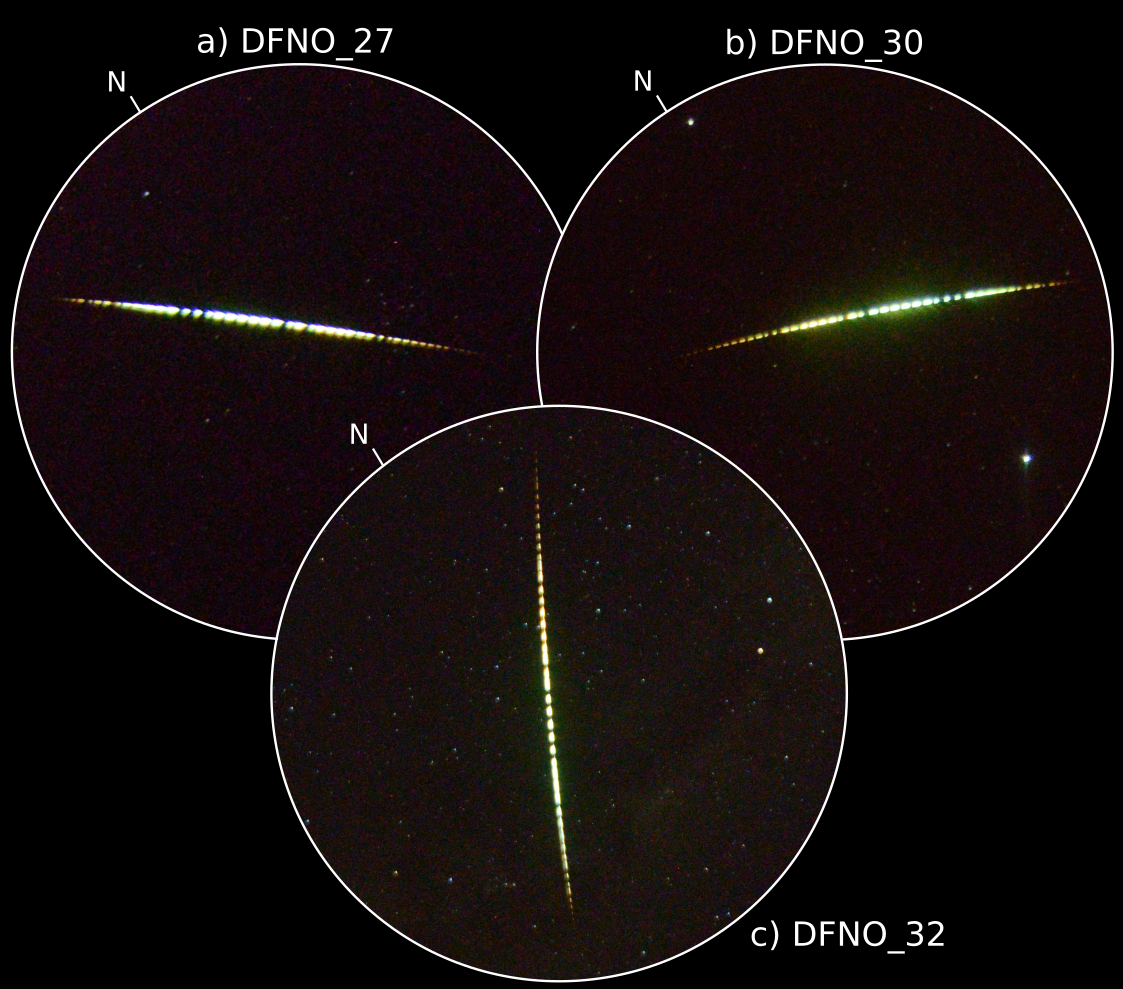}
    \caption{\textit{DN160410\_03} fireball as seen from three DFN stations in South Australia, starting at 13:09:02.526 UTC. Calibration with background stars determines azimuth and elevation of trajectory points.}
    \label{fig:imgs}
\end{figure}

\begin{figure}
    \centering
    \includegraphics[width=\textwidth]{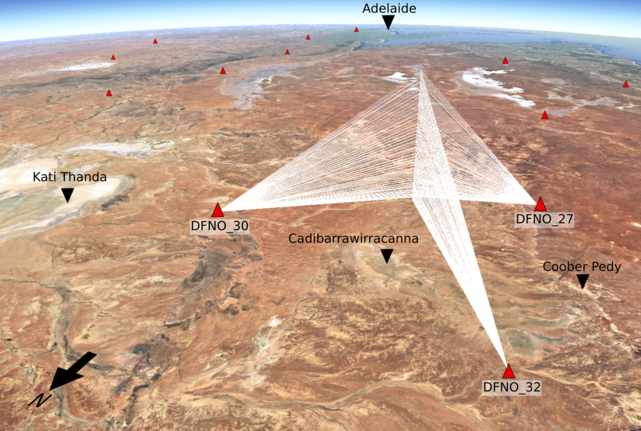}
    \caption{Configuration of \textit{DN160410\_03} observations. White observation rays correspond to the start and end points of the trajectory dashes in Figure \ref{fig:imgs}. (Google Earth image credit: Landsat/Copernicus).}
    \label{fig:rays}
\end{figure}

On the 10th of April 2016, at 13:09:02.526 UTC, a `typical' fireball was observed by three DFN observatories over central South Australia, near lake Cadibarrawirracanna (event \textit{DN160410\_03}, Figure \ref{fig:imgs}). It is an ideal case to analyse as it was nearly equidistant to all cameras, with the angle of observing planes at $46^\circ/52^\circ/80^\circ$ (Figure \ref{fig:rays}), with 88 of the 91 total observations made (from identifying the starts and ends of the trajectory dashes) were visible in all three still images (Figure \ref{fig:imgs}). There is little observable fragmentation in the still images and no major peaks noticeable in the light curve which is regular and symmetric. The method used by the DFN to calculate the luminosity of an event is only applicable when an event does not saturate the sensor, which was unfortunately the case for the other two viewpoints, DFNO\_27 and DFNO\_32. 

\begin{figure}
    \centering
    \includegraphics[width=\textwidth]{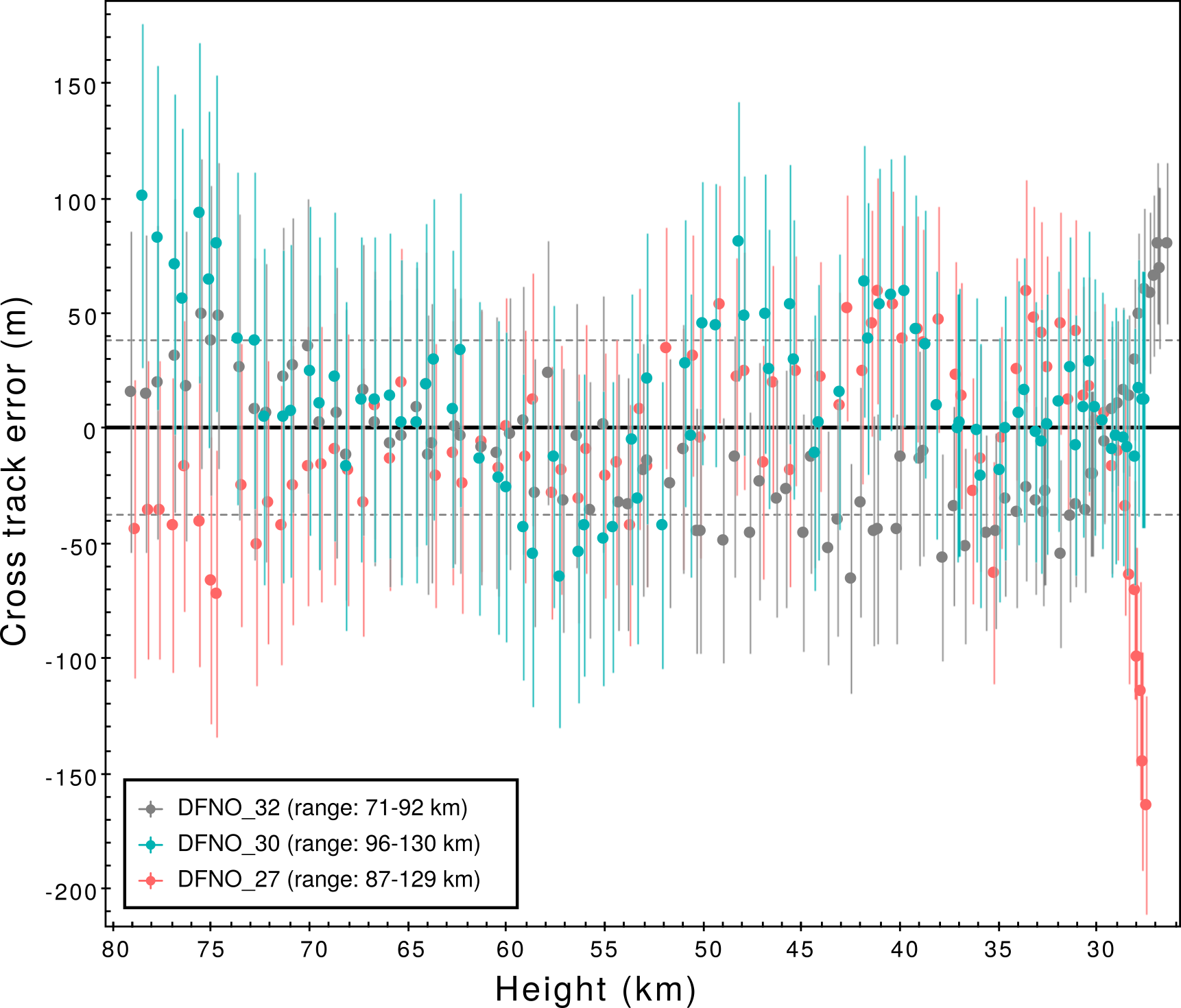}
    \caption{Cross track residuals for individual station observations of \textit{DN160410\_03} to a straight line trajectory fit in an inertial (ECI) reference frame. Range values in legend are the minimum and maximum distances of a station to the fireball trajectory. Error bars on observations are $1\sigma$ astrometric errors propagated by the corresponding range.}
    \label{fig:ba_xtrack}
\end{figure}

As with event \textit{DN151212\_03}, a straight line least squares (SLLS) triangulation of this event was calculated. Initially for the entire trajectory (Figure \ref{fig:ba_xtrack}), with resulting parameters determined using the EKS given in Table \ref{tab:ba_slls}. Although this event is steeper and significantly shorter, gravity still contributes a 105\,m  downward component over the 58\,km long trajectory and Earth rotation an apparent 1.9\,km lateral deflection to an observer on the ground. The ECI and ECEF entry radiants show a $1.74^\circ$ separation (Table \ref{tab:ba_slls}).
\begin{table}[]
    \centering
    \begin{tabular}{l|c|c}
         \textit{DN160410\_03} (full)& ECI & ECEF \\\hline\hline
         entry radiant -- RA ($\,^\circ$) &$161.91\pm0.01$&$163.49\pm0.03$ \\\hline
          entry radiant -- DEC  ($\,^\circ$) &$-4.52\pm0.01$&$-4.64\pm0.01$ \\\hline
          initial height (km)& $79.1\pm0.05$
                        & $79.1\pm0.02$\\\hline
         initial velocity (km s$^{-1}$) & $15.2\pm0.1$
                          & $15.2\pm0.1$\\ \hline
         initial mass (kg)& $1.6\pm0.7$ 
                          & $1.5\pm0.6$\\ \hline 
        entry slope, $\gamma_e$ ($^\circ$)  & $64.3$
                     & $64.8$\\\hline
         final height (km)& $26.7\pm0.07$
                      & $26.6\pm0.05$\\ \hline
         final velocity (km s$^{-1}$) & $4.0\pm0.4$
                        & $4.0\pm0.7$\\ \hline
         final mass (kg)& $0.05\pm0.01$ 
                          & $0.06\pm0.01$\\ \hline 
    \end{tabular}
    \caption{Straight line least squares (SLLS) trajectory triangulated in either an inertial (ECI) or non-inertial (ECEF) reference frame for all observations of event  \textit{DN160410\_03}. Trajectory characteristics (height, velocity, mass) are estimated using an extended Kalman smoother in one dimension on these straight line data. The angular separation between the two radiants is $1.74^\circ$.}
    \label{tab:ba_slls}
\end{table}

Despite the apparently reasonable fit of the straight line to the entire trajectory in this case, we once again isolate the observations above 50 km and re-triangulate this upper dataset. The ECI entry radiant changed by a not insignificant $0.16^\circ$ (Table \ref{tab:ba_slls_tops}). 
\begin{table}[]
    \centering
    \begin{tabular}{l|c|c}
         \textit{DN160410\_03} ($>50\,km$) & ECI & ECEF \\ \hline\hline
         entry radiant -- RA ($\,^\circ$) &$161.86\pm0.01$&$163.41\pm$ \\\hline
          entry radiant -- DEC  ($\,^\circ$) &$-4.54\pm0.01$&$-4.56\pm0.01$ \\\hline
          initial height (km) & $79.13\pm0.01$
                        & $79.12\pm0.01$\\\hline
         initial velocity (km s$^{-1}$) &  $15.22\pm0.06$
                          & $15.18\pm0.02$\\ \hline
         entry slope, $\gamma_e$ ($^\circ$) & 64.2
                     & 64.7\\ \hline
    \end{tabular}
    \caption{ Straight line least squares (SLLS) trajectory triangulated in either an inertial (ECI) or non-inertial (ECEF) reference frame for observations of event \textit{DN160410\_03} above 50 km only. Trajectory characteristics (height, velocity, mass) are estimated using an extended Kalman smoother in one dimension on these straight line data. The angular separation between the two radiants is $1.55^\circ$.}
    \label{tab:ba_slls_tops}
\end{table}

\begin{figure}
    \centering
    \includegraphics[width=\textwidth]{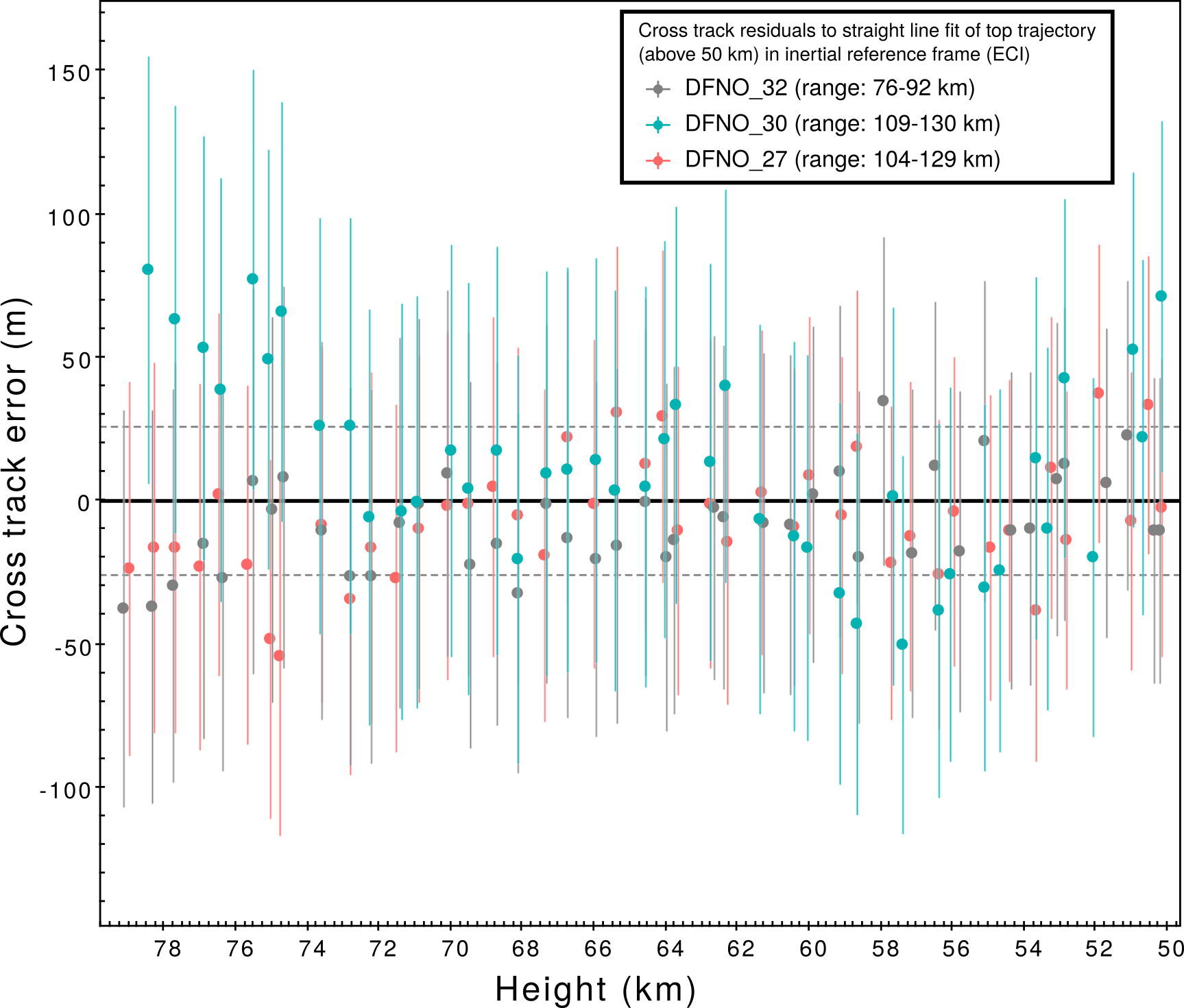}
    \caption{Cross track residuals for the upper section the \textit{DN160410\_03} trajectory to a straight line fit in an inertial (red; ECI) and non-inertial (blue; ECEF) reference frame. Only observations of the fireball above $50\,km$ were used. Error bars on observations are $1\sigma$ astrometric errors propagated by the corresponding range. }
    \label{fig:ba_xtrack_tops}
\end{figure}
\begin{figure}
    \centering
    \includegraphics[width=\textwidth]{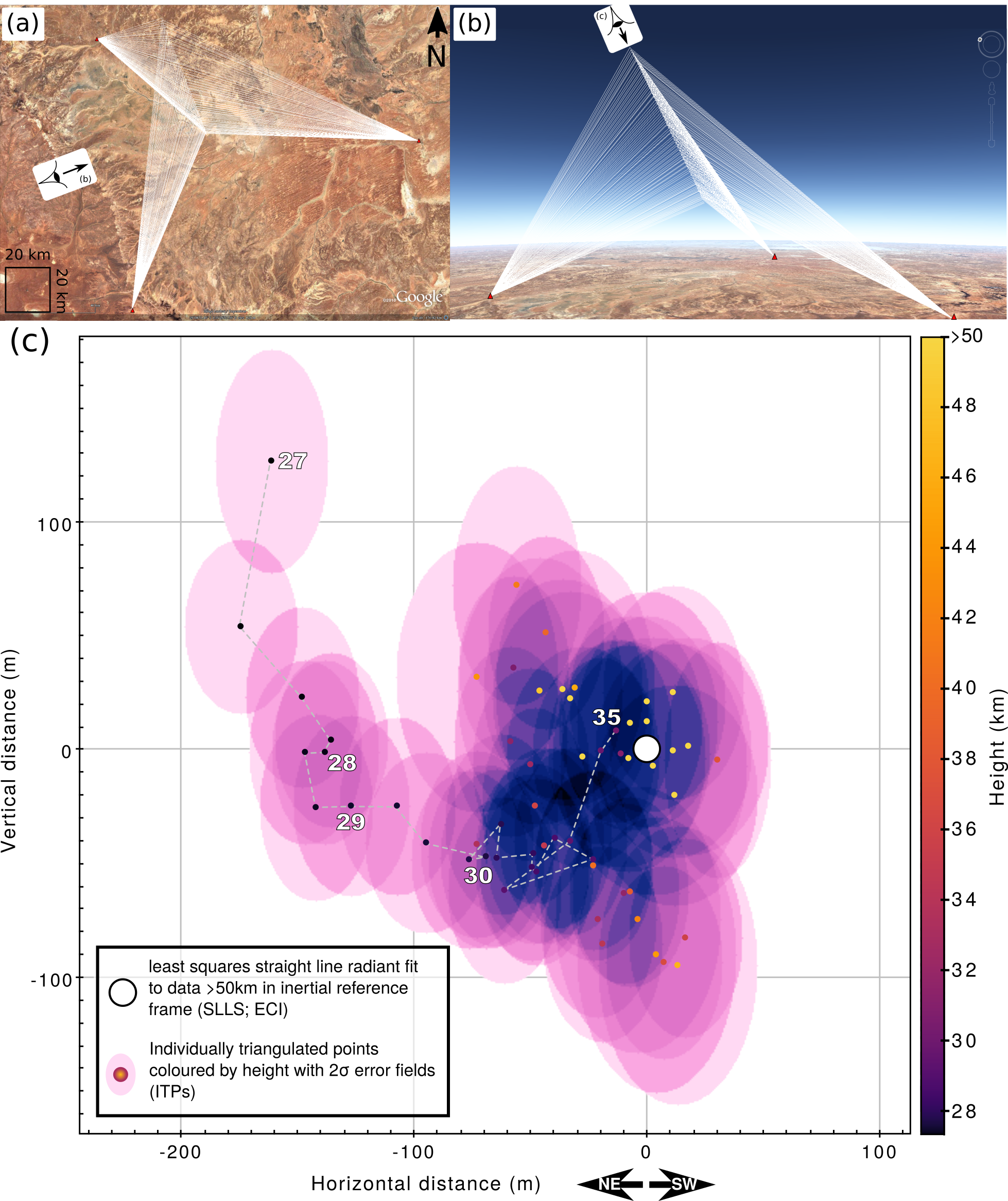}
    \caption{\textit{DN160410\_03} -- ``Down-line" view (c) as seen by an observer looking down the ECI straight line radiant (illustrated by (a)--(b)) calculated using the top section of the trajectory (points $>50\,km$; see Figure \ref{fig:ba_xtrack_tops}). This results in the ECI trajectory stacking to a single point (white) whereas the individually triangulated positions (ITPs; coloured points) are projected onto the viewing plane. This plane is normal to the straight line trajectory with the x-axis aligned with the Earth horizontal, and inclined from true vertical by the cosine of the trajectory slope. 
    From this down-line view the ITPs help to illustrate the true non-linearity of the path taken by the meteoroid. (Google Earth image credit: Landsat/Copernicus/CNES/Airbus).}
    \label{fig:ba_downline}
\end{figure}

With this new entry radiant, we can once again project the ITPs onto the plane normal to it, allowing us to observe how the meteoroid positions track in the lower section of the trajectory (Figure \ref{fig:ba_downline}).
Despite this event being a more typical example, with acceptable observational residuals, there is still a not insignificant lateral trend to the end meteoroid positions as shown by the ITPs in this figure. 

\subsubsection{Summary of straight line comparisons}
Event \textit{DN151212\_03} may be considered unique in its duration, and its non-linear flight path with up to 2.1\,km of lateral deviation an anomaly, but in performing a similar analysis to the more typical event \textit{DN160410\_03} we are still able to notice a distinct pattern/wander to the end of the trajectory. These deviations from a straight line cannot be accounted for by gravity, and Earth rotation effects are removed when an inertial reference frame is used. 
It is clear that there are real limitations to the straight line assumption and it is best to consider modelling fireball trajectories without any pre-defined assumptions; allow the raw observations to be the sole influences on the data. 
To achieve this we can apply the single body equations 
in three dimensions to the particle filter methodology described by \citet{Sansom2017}. This will utilise raw astrometric observations to resolve meteoroid position estimates.

\section{Particle Filter Modelling Using Three Dimensional Meteoroid Flight and Luminosity}
\label{sec:pfmethods}

The iterative Monte Carlo technique of the particle filter allows a broad range of trajectory parameters (including densities, shapes and ablation parameters) to be explored, and favourable values to be identified, in a more robust way than a brute force least squares approach. 
A set of tracer particles are propagated through the motion and luminous equations, and their weightings evaluated at each time step according to their closeness to available observational data. A detailed description of applying particle filters to meteoroid trajectories is presented in \citet{Sansom2017}. 
Although \citet{Sansom2017} apply the single body equations as a model, the adaptive approach uses appropriate covariances to incorporate, to some extent, unmodelled processes such as fragmentation.

Particle filters fall within the class of Bayesian state-space methods which use a vector, $\mathbf{x}$, to represent the  \emph{state} of a system. In meteoroid trajectory analysis this includes the motion parameters (position and velocity) as well as other trajectory variables.

To use a three dimensional model for flight, we divide positions and velocities into their $x, y, z$ components in geocentric coordinates.  
Incorporating the luminous efficiency into the state vector allows luminosity values to be calculated. Equation \eqref{eqn:state} represents the meteoroid state and encapsulates the knowledge of the meteoroid system at a given time $t_k$.

\begin{equation}\label{eqn:state}
\mathbf{x}_k =  \left[ \begin{array}{c}
\ell_{x;k} \\
\ell_{y;k}\\
\ell_{z;k}\\
v_{x;k}\\
v_{y;k}\\
v_{z;k}\\
m_k \\
\kappa_k \\
\sigma_k\\
\tau_k
\end{array} \right] 
\begin{array}{l}
\text{position in X} \\
\text{position in Y} \\
\text{position in Z} \\
\text{velocity in X} \\
\text{velocity in Y} \\
\text{velocity in Z} \\
\text{mass}  \\
\text{shape-density parameter} \\
\text{ablation parameter}\\
\text{luminous efficiency parameter}.
\end{array}
\end{equation}

The shape-density and ablation trajectory parameters are given by 
\begin{equation*}
\kappa = \frac{1}{2}\frac{c_d A }{\rho_m^{2/3}} \qquad \text{  and  } \qquad \sigma =  \frac{c_h }{c_dH^*}  ,
\end{equation*}
where $c_d$ and $c_h$ the drag and heat transfer coefficients respectively, $A$ is the shape parameter as described by \citet{Bronshten1983}, 
$\rho_m$ the bulk density of the meteoroid, and $H^*$ the enthalpy of sublimation. 

This state is determined by assessing the conditional probability density function
$p(\mathbf{x}_k|\mathbf{z}_{0:k})$ 
given an observation $\mathbf{z}_{k}$  of the system at time $t_k$ ($\mathbf{z}_{0:k}$ therefore being the history of all observations from time $t_0$ through to time $t_k$).

This is achieved through the three state-space equations:
\begin{flalign}
&\text{(i)   The state prior, }&&p(\mathbf{x}_0),&\label{eqn:prior}
\intertext{which is the probability density function that encapsulates prior knowledge of the state of the system and initialises the recursion.}
&\text{(ii)  The process equation, }& & \mathbf{x}_{k+1} = f(\mathbf{x}_{k}) + \mathbf{u}_k,\label{eqn:proc}
\intertext{defines the evolution of the state in discrete time, with process noise $\mathbf{u}_k$.}
&\text{(iii) The measurement equation, }&&\mathbf{z}_k =h( \mathbf{x}_k) +  \mathbf{w}_k,&\label{eqn:meas}
\end{flalign}
uses the measurement function $h(\mathbf{x}_k) $ to correlate the state of the meteoroid to the given azimuth and elevation measurements from camera observatories. Observation noise, $\mathbf{w}_k$, is assumed to be Gaussian with a mean of zero and covariance $\mathbf{R}_k$ in degrees ($\mathbf{R}_k$ represents observational error). Further explanation of the measurement function are detailed in Section \ref{ch05:sec_init}.

Although fireball observations are made in discrete time, modelling the meteoroid dynamics is more appropriate using continuous model equations. 
Continuous-time differential state equations $(f_c(\mathbf{x}))$
may be integrated in order to attain the form needed for the process equation \eqref{eqn:proc}:
\begin{equation}\label{eqn:continuous}
\mathbf{x}_{k+1} =  \int_{t_k}^{t_{k+1}}f_c(\mathbf{x}) \mathop{dt} + \mathbf{u}_k.
\end{equation}

Although $f_c(\mathbf{x}) $, using the single body equations, is non-linear, the discrete-time process noise, $\mathbf{u}_k$, can be closely approximated by Gaussian noise with zero mean and covariance $\mathbf{Q}_k$ ($\mathbf{Q}_k$ corresponds to how well the process equation represents the true system).

A particle filter is very flexible and requires no constraints on the linearity of state equations, nor the noise distributions \citep{Ristic2004}.  This is due to there being no single representation of the state prior, rather a set of $N_s$ weighted particles are used to represent the distribution. 

Each $i$th particle can be represented at any time $t_k$ by its state, $\mathbf{x}_{k}^i$, and weight, $w_k^i$ as:
\begin{equation}
\{\mathbf{x}_{k}^i, w_k^i \} \quad i = 1, ..., N_s,
\end{equation}
with weights normalised as:
\begin{equation}\label{eqn:sum1}
\sum_i^{N_s}  w_k^i = 1.
\end{equation}
Particle weights are evaluated according to how well a particle's state represents the available observational data. 
The weighted mean of the distribution, $\hat{\mathbf{x}}_k$, can be approximated at any time $t_k$  as:
\begin{equation}\label{eqn:mean}
\hat{\mathbf{x}}_k= \sum_i^{N_s} w_k^i\mathbf{x}_k^i,
\end{equation}
with covariance 
\begin{equation}\label{eqn:var}
Cov( \mathbf{x}_k) = \sum_i^{N_s}w_k^i (\mathbf{x}_k^i - \hat{\mathbf{x}}_k)(\mathbf{x}_k^i - \hat{\mathbf{x}}_k)^T.
\end{equation}

There are three steps in running a particle filter, similar to other Bayesian filtering methods: \textit{initialisation, prediction, update}. 
\citet{Sansom2017} provides a detailed methodology for a one dimensional trajectory model.  
Here we will outline the variations required to allow a particle filter to be performed in three dimensions and incorporate absolute visual magnitude observations. 

\subsection{Initialisation using point-wise triangulation}\label{ch05:sec_init}

An initial set of particles is required that best represents the state prior \eqref{eqn:prior} of the meteoroid system; initialisation in 3D requires an approximate start location.
As the full data set is available at the time of executing the particle filter, the initial position and velocity components may be more accurately estimated from the observational data. The instantaneous meteoroid position for a given time step can be evaluated using \textit{point-wise triangulation} (see Section \ref{sec:pw}).
Performing a point-wise triangulation on the first handful of multi-station observations produces a set of individually triangulated positions (ITPs) from which the instantaneous velocity of the meteoroid can be determined  -- simply taking the difference in positions with time: $\mathbf{v}_{k} = \cfrac{d\Bell_{k:k+1}}{dt_{k:k+1}}$.
Due to the inherent scatter in the ITPs and therefore velocities, $v_0$ may be reasonably well approximated by assuming constant deceleration between the first few multi-station observations and $t_0$:
\begin{equation}
\label{eqn:line}
\mathbf{v}_0 =\mathbf{v}_{m} -  \cfrac{d\mathbf{v}}{dt}\times t_{m},
\end{equation}
where $t_m$ is the relative time of the first available \textit{multi-station} observations and the value of $\mathbf{v}_m$ and $\cfrac{d\mathbf{v}}{dt}$ are determined by a linear least squares fit to the scattered velocities.

If the first ITP is at $t_0$, its position can be used to initialise the first three components of the state vector \eqref{eqn:state}. 
If $t_m \neq t_0$ (multi-station data is not available at $t_0$), an initial position may then be roughly approximated by rearranging and integrating \eqref{eqn:line} with respect to time:
\begin{equation}
\mathbf{\Bell_0} = \Bell_m - \mathbf{v}_{0} t_m - \cfrac{1}{2} \cfrac{d\mathbf{v}}{dt}\times{t_m}^{2}.
\end{equation}

The initialisation of particle state parameters for position and velocity at $t_0$ is then drawn from a Gaussian distribution shown by  
\begin{equation}\label{eqn:lv_init}
\Bell_0^i =  \mathcal{N}(\Bell_0; \mathbf{P}_{\Bell;0}) \qquad
\mathbf{v}_0^i =  \mathcal{N}(\mathbf{v}_0; \mathbf{P}_{\mathbf{v};0}) \qquad
i = 1,...,N_s
\end{equation}
where mean values of the $\Bell_0$ and $\mathbf{v}_0$ vectors are calculated as described above, and covariance values, $\mathbf{P}_0$, are determined by the uncertainty in this least squares fit and may vary for each directional component. 

Possible original values for mass, $\kappa$ and $\sigma$ can be randomised within theoretical bounds (see Table 1 of \citealt{Sansom2017}). A similar concept can be applied to the luminous efficiency; here we randomise within the range $0.01\%<\tau<10\%$ after \citet{Ceplecha2005} and \citet{Ceplecha1998}.
All particles are initially weighted equally as $w_0^i = \cfrac{1}{N_s}$. 

\subsection{Filter Prediction Using Three Dimensional State Equations}
Recursion commences after initialisation, beginning with a forward prediction of particles from $t_{k}$ to $t_{k+1}$ by the process equation \eqref{eqn:proc}.

The change in trajectory parameters  $\kappa$, $\sigma$ and $\tau$ with time is not well known and at this stage is assumed to be nil (see discussion related to Equation \eqref{eqn:Qc}):
\begin{align} 	  	\label{eqn:params}
\frac{d\kappa}{dt}    = 	\frac{d\sigma}{dt}   =\frac{d\tau}{dt}   = 0  .
\end{align}

In order to analyse the full trajectory in 3D, the differential equations of motion must be split into their vector components:

\begin{subequations} 
	\label{eqn:components}
	\begin{align}
	\cfrac{d\Bell}{dt}& =\mathbf{v}\\
	\cfrac{d\mathbf{v}}{dt}&= -\kappa \rho_a m^{(\mu-1)}  ||\mathbf{v}||
	\mathbf{v}+ 	\mathbf{g}\label{eqn:dyn_v2}\\
	\cfrac{dm}{dt}&=-\kappa \sigma \rho_a  m ^{\mu}   ||\mathbf{v}|| ^3,\label{eqn:dyn_m2}
	\end{align}
\end{subequations}    
where $\Bell$ and $\mathbf{v} $ are the position and velocity vectors, $\mathbf{g}$ is the local gravitational acceleration vector, and $||\mathbf{v}||$ is the magnitude of the velocity.  
$\mu$ is the shape change parameter, representing the rotation of the body, here assumed to be $2/3$, representing spin rapid enough for ablation to be uniform across the entire surface \citep{Bronshten1983}. 
Atmospheric densities, $\rho_a$, can be calculated using the NRLMSISE-00 atmospheric model \citep{Picone2002}.

This gives the continuous-time state equation for a meteoroid travelling through the atmosphere in 3D as:
\begin{equation}\label{eqn:fc}
f_c(\mathbf{x}) = \left[
\cfrac{dl_x}{dt}, 
\cfrac{dl_y}{dt}, 
\cfrac{dl_z}{dt}, 
\cfrac{dv_x}{dt}, 
\cfrac{dv_y}{dt}, 
\cfrac{dv_z}{dt},
\cfrac{dm}{dt},
\cfrac{d\kappa}{dt},
\cfrac{d\sigma}{dt},
\cfrac{d\tau}{dt}
\right].
\end{equation}
with the continuous-time Gaussian process noise $\mathbf{u}_c$ of zero mean and covariance  $\mathbf{Q}_c$. The discrete-time process noise covariance, $\mathbf{Q}_k$ can be approximated as   
\begin{equation}
\label{eqn:Q}
\mathbf{Q}_k = \int_{t_k}^{t_{k+1}}e^{Ft} \, \mathbf{Q}_c \, e^{F^T t} \mathop{dt}
\end{equation}
using the linearised form of the process equation, $F  = \cfrac{\partial  f_c(\mathbf{x})}{\partial \mathbf{x}}$ \citep{Grewal1993}. In the 3D filter, we use 
\begin{equation}
\begin{aligned}
\mathbf{Q}_c = 	diag & [(0\,\text{\tiny$m\,s^{-1}$}), (0\,\text{\tiny$m\,s^{-1}$}), (0\,\text{\tiny$m\,s^{-1}$}),\\
& (75\,\text{\tiny$m\,s^{-2}$}), (75\,\text{\tiny$m\,s^{-2}$}), (75\,\text{\tiny$m\,s^{-2}$}),\\
&(0.8\times m_k\,\,\text{\tiny$kg\,s^{-1}$}), (10^{-3}\,\text{\tiny$m^2\,kg^{-2/3}\,s^{-1}$}), \\
&(10^{-4} \, \text{\tiny$s\,km^{-2}$}), (0.001 \, \text{\tiny\%})]^2,
\label{eqn:Qc}
\end{aligned}
\end{equation}
where each element along this square matrix diagonal represents the uncertainty of each differential model equation in \eqref{eqn:fc}. 
That is, the uncertainty in position and velocity components are introduced through noise in the acceleration model \eqref{eqn:dyn_v2}, therefore allowing the variance of $d\Bell/dt=0\,m\,s^{-1}$. The process model for $dm/dt$ is not able to fully represent the change of mass due to fragmentation; the process noise is therefore set as a relatively high percentage of the existing mass. 
Although the trajectory parameters $\kappa$, $\sigma$ and $\tau$ are currently assumed to be constant \eqref{eqn:params}, this is not entirely true; process noise is therefore attributed to allow small variations to these parameters throughout the trajectory  \eqref{eqn:Qc}.  

The discrete process noise, $\mathbf{Q}_k$, is then calculated from Equation  \eqref{eqn:Q} at every time step along the trajectory.

\subsection{Line-Of-Sight Measurement Update}\label{sec:meas}

Images taken by each observatory show a discontinuous streak across a star background. The Desert Fireball Network uses the modulation of a liquid crystal shutter within the lens of each camera to encode a unique time sequence into the fireball's path \citep{howie2017deb}. By comparing the position of the start and end of each fireball segment with the background stars, the azimuth and elevation of each time encoded data point can be determined \citep{2018arXiv180302557D}.

The astrometric observations of the fireball, $\mathbf{z}_k$, are a series of angular measurements.
The measurement function in Equation \eqref{eqn:meas} extracts the position vector ($\Bell$) from the state which will be compared to these observations and performs the transformation required. 
Within this function, $\Bell_k^i$ is converted from geocentric cartesian coordinates to a calculated line-of-sight azimuth and elevation with respect to each observatory. At each $t_k$, this conversion is required for each station that made an observation.  
The cartesian vector between each $n$ observatory and the particle position, is rotated into local observatory-centred coordinates (East, North, Up; $\left[\mathbf{\hat{a}}_k^{i;n}\right]_{ENU}$) before subdividing it into its altitude and elevation components:

\begin{align}
az_{k}^{i;n} &= \arctan2\left(\left[\mathbf{\hat{a}}_k^{i;n}\right]_E\, ,\,\,\left[\mathbf{\hat{a}}_k^{i;n}\right]_N \right)\qquad(mod\,2\pi)\label{eqn:az}\\
el_{k}^{i;n} &= \arcsin\left(\left[\mathbf{\hat{a}}_k^{i;n}\right]_U\right)\label{eqn:el}.
\end{align}
For consistency in calculated and true angular measurements, the azimuth value is expressed within the 0 to $2\pi$ radian range. As an azimuth value of $0$ radians is congruent with that of $2\pi$ radians, a modulo operation is included in \eqref{eqn:az}.

The result of the measurement function, $\mathbf{\hat{z}}_{k}^{i} $ is the predicted line-of-sight unit vectors for a given particle $i$ in azimuth and elevation from all observatories and can be summarised by: 
\begin{equation}
\mathbf{\hat{z}}_{k}^{i} = \left[az_{k}^{i;1}, el_{k}^{i;1}, az_{k}^{i;2}, el_{k}^{i;2},...\right]. 
\end{equation}
A multivariate Gaussian probability is then used to calculate the \textit{position} weighting of a particle:
\begin{equation}\label{eqn:pos_weight}
\left[\tilde{w}_k^i\right]_{pos} =	\left(2 \pi^{\frac{N_s}{2}}|\mathbf{R}_k|^{\frac{1}{2}}\right)^{-1} e^{-\frac{1}{2}\left[\mathbf{\hat{z}}_k-\mathbf{z}_k\right]^T \mathbf{R}_k^{-1}\left[\mathbf{\hat{z}}_k^i-\mathbf{z}_k\right]} ,
\end{equation}
where $|\mathbf{R}_k|$ is the determinant of the observation noise covariance matrix containing azimuth and elevation errors pertaining to each observatory:
\begin{equation}
\mathbf{R}_k = diag[Var(az^{1}) , Var(el^{1}), Var(az^{2}), Var(el^{2}), ...].
\end{equation}
The observational uncertainties in both azimuth and elevation are linked to the accuracy of picking the start and end points of modulated segments in the fireball image, their calibration and the shutter response time. For all-sky images captured using fish eye lenses, the accuracy in azimuth is much greater than in elevation. Although the DFN observations are syncronised in time, this is not required by the 3D particle filter; only muti-station observations which include absolute timing data are needed.

\subsection{Luminosity measurement update}
As well as considering the line-of-sight observations, the calculated absolute visual magnitude observations, $M_v^{obs}$, may also be used to constrain mass loss estimates. Observed luminosities can be obtained from the long exposure images by doing aperture photometry on each shutter break.
These measurements are then converted to apparent magnitudes using the stars, accounting for the different exposure times.
Apparent magnitudes are finally turned into absolute magnitudes ($M_v^{obs}$) by doing a distance correction using the basic trajectory solution given by the SLLS.
A combination of Equations \eqref{eqn:I} and \eqref{eqn:Iv} are used to calculate the predicted visual magnitude for each particle, $M_v^i$ between $t_k$ and $t_{k+1}$. 
The \textit{luminous} weighting for each particle, $\left[\tilde{w}_{k}^i\right]_{lum}$ can then be obtained by evaluating the 1D Gaussian probability distribution function 
\begin{equation}
        \left[\tilde{w}_{k}^i\right]_{lum} = \frac{1}{\sqrt{2R_k\pi}}e^{-\frac{(M_v^{obs}-M_v^i)^2}{2R_k}}
\end{equation}

where $R_k$ here is the uncertainty in the observed $M_v$ values. This can include errors introduced in the calibration process that is usually required to convert arbitrary brightness units to distance-normalised, absolute visual magnitudes. 

The overall weighting of each particle including both line-of-sight and absolute magnitude observations can then be calculated as the product of normalised position and luminous weightings:

\begin{equation}\label{eq:weight}
    \tilde{w}_{k}^i = \left[\tilde{w}_{k}^i\right]_{pos} \left[\tilde{w}_{k}^i\right]_{lum},
\end{equation}
which can then be normalised
\begin{equation}
    w_{k}^i = \frac{\tilde{w}_{k}^i}{\sum_j^{N_s}\tilde{w}_{k}^j}.
\end{equation}

\subsection{Results of the 3D particle filter}

\begin{figure}
	\centering
	\includegraphics[width=\textwidth]{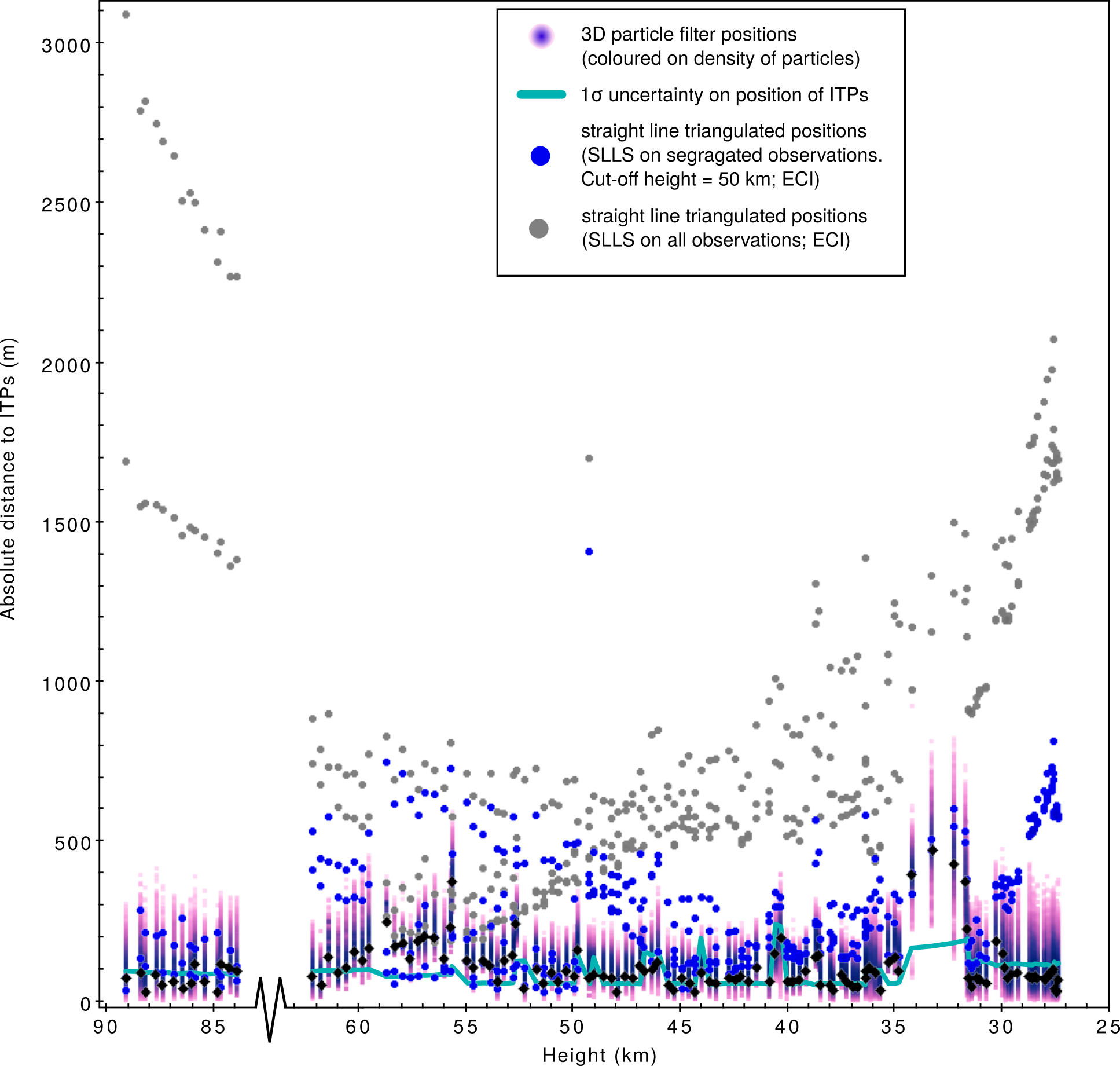}
	\caption{The absolute distance between individually triangulated positions (ITPs; $y=0$ with variances in green) and the estimated position of the \textit{DN151212\_03} meteoroid using different methods of modelling a meteoroid trajectory: a straight line least squares approximation (SLLS) fitted to the entire suite of observations in an ECI reference frame (grey), a SLLS fitted to the upper (above $50\,km$) and lower (below $50\,km$) segments of the trajectory separately (blue; see Section \ref{sec:slls}) and the results of a 3D particle filter (weighted mean positions in black). 
	The gap between $62-84\,km$ corresponds to the time between exposures. The final $1.22\,s$ ($\sim800\,m$ height) was only observed by a single observatory and no individually triangulated position could be calculated as a reference. 
	}
	\label{fig:bv_diffs}
\end{figure}

\begin{figure}
    \centering
    \includegraphics[width=\textwidth]{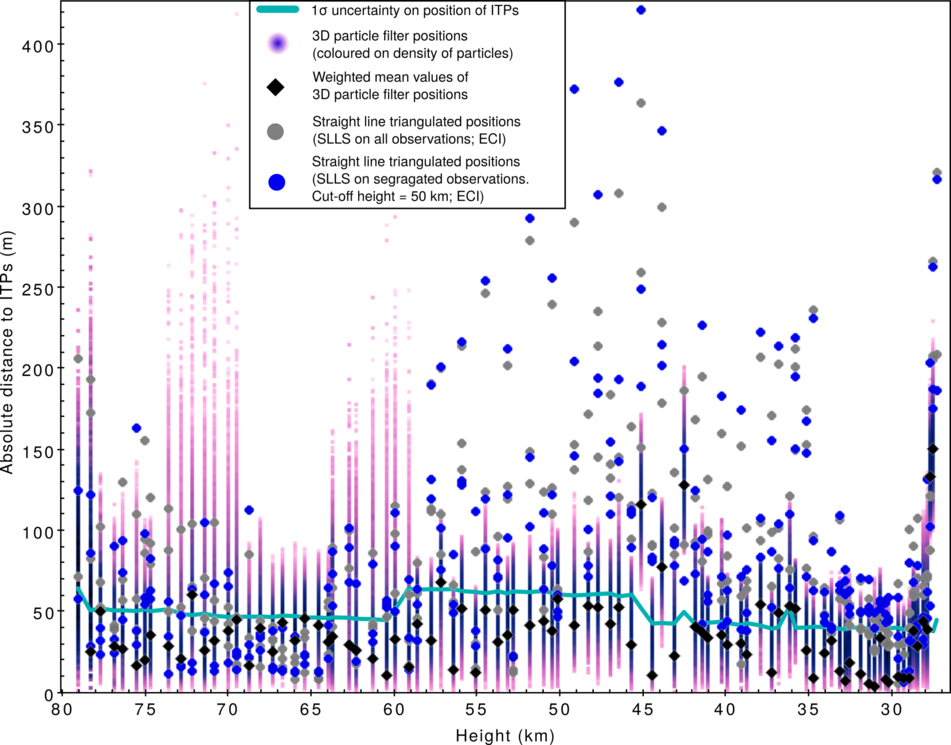}
    \caption{The absolute distance between individually triangulated positions (ITPs; $y=0$ with normalised variances in green) and the estimated position of the \textit{DN160410\_03} meteoroid using different methods of modelling a meteoroid trajectory: a straight line least squares approximation (SLLS) fitted to the entire suite of observations in an ECI reference frame (grey), a SLLS fitted to the upper (above $50\,km$) and lower (below $50\,km$) segments of the trajectory separately (blue; see Section \ref{sec:slls}) and the results of a 3D particle filter (weighted mean positions in black). }
    \label{fig:ba_diffs}
\end{figure}

A 3D particle filter was performed on the two test cases using $N_s=100,000$. 
The distance between the ITPs and all predicted particle positions for event \textit{DN151212\_03} are shown\footnote{In order to graphically represent such a large output file (number of particles plotted $= N_s\times k$) we made use of TOPCAT table processing software which is an open source library for manipulating large tabular data \citep{taylor2005topcat}} in Figure \ref{fig:bv_diffs} and for event \textit{DN160410\_03} in Figure \ref{fig:ba_diffs}. 
The weighted mean residuals, as calculated by Equation \eqref{eqn:mean}, are marked in black.

ITPs may give us a reasonable indication of meteoroid position, but are sensitive to observational geometry and error. Despite using the ITPs as reference positions for these figures, the 3D particle filter is weighting estimates based on the raw observations. The history information inherent in the particle `cloud' provides a certain inertia that prevents unrealistic changes to the overall mean when unfavourable observations are made.
Though there is still a flexibility to the estimates that allows to an extent for unmodelled factors (fragmentation etc. not included in the single body equations) to be incorporated, as model uncertainty is considered in the process noise covariance (Equation \eqref{eqn:Qc}).

\subsubsection{Case 1: DN151212\_03}

No absolute brightness data were acquired for this event as the fireball saturated the sensors. The particle filter was still able to calculate theoretical values for $M_v$, though only normalised values of $\left[\tilde{w}_k^i\right]_{pos}$  (Equation \ref{eqn:pos_weight}) were used to determine particle weightings for this case.

\begin{figure}
    \centering
    \includegraphics[width=\textwidth]{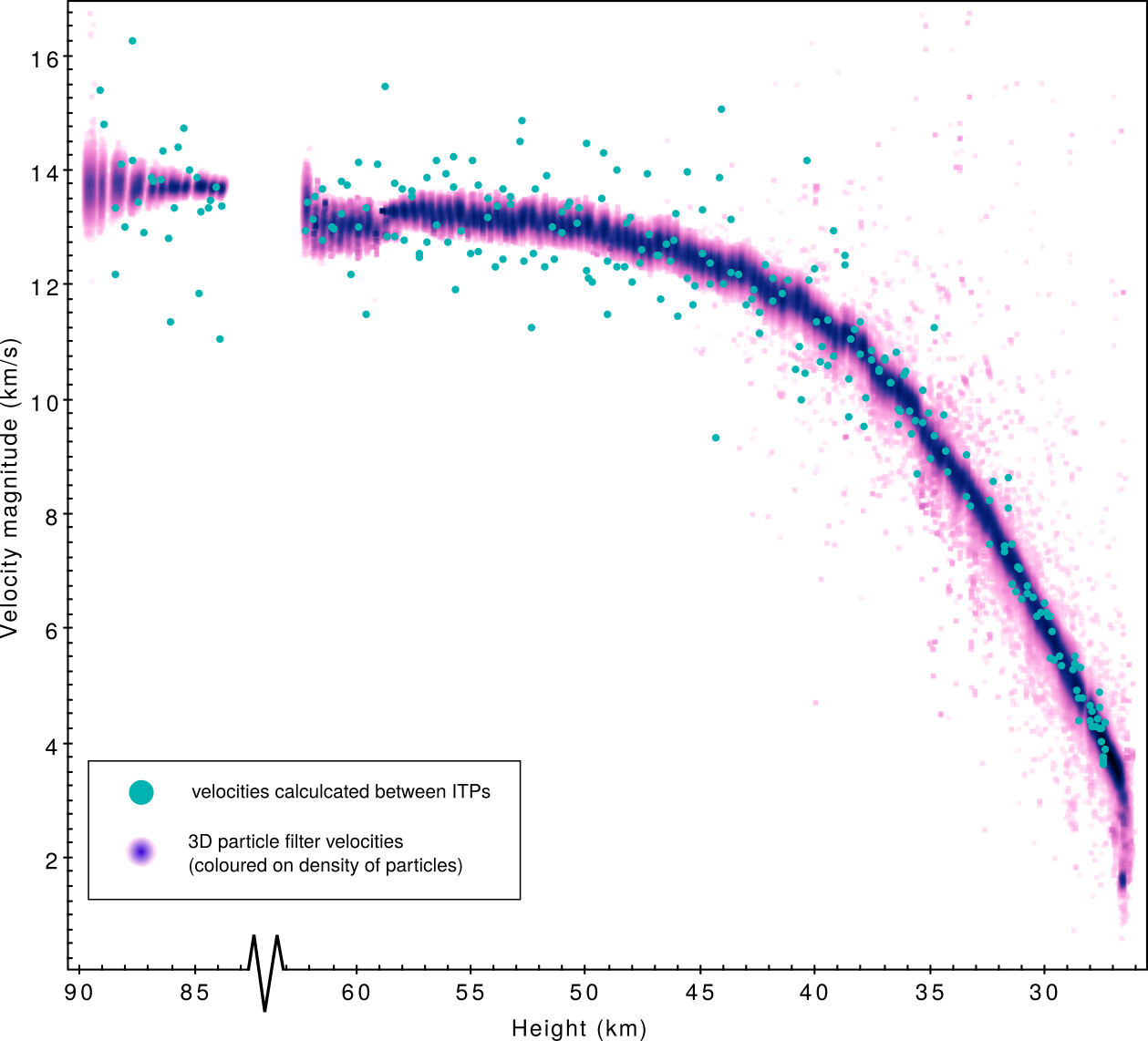}
    \caption{Magnitude of the velocity vector as calculated by the change in ITP positions with time (green) and as estimated by the 3D particle filter. The gap between 62-84 km corresponds to the time between exposures.} 
    \label{fig:bv_vels}
\end{figure}

The maximum deviation of any weighted mean to its corresponding calculated observed position for \textit{DN151212\_03} is $470\, m$, with over half within $80\,m$. The higher mean values between $34$ to $32\,km$ could be related to increased uncertainty in the ITPs for these observations (Figure \ref{fig:bv_diffs}), or could be indicative of an unmodelled cause.
The large gap in Figure \ref{fig:bv_diffs} between $62-84\,km$ corresponds to the time between exposures. 
For this event, the final $1.22\,s$ (seven observation times) were only visible from one camera (Figure \ref{fig:151212_39}). 
The 3D particle filter still estimates positions with single station observations, but the mean final state estimate at $t_f =21.14\,s$ has slightly higher uncertainties because of this.
Particles are not shown in Figure \ref{fig:bv_diffs} for this final $760\,m$ as no individually triangulated position could be calculated as a reference. The exploration of velocity state-space by the particles can be seen in Figure \ref{fig:bv_vels}. Final state estimates are given in Table \ref{tab:3dpf_tf}.

\begin{table}[]
    \centering
    \begin{tabular}{l|c|c}
         final state values for: &\textit{DN151212\_03}  & \textit{DN160410\_03}\\ \hline\hline
         $t_f$ (seconds since start) & 21.14 & 4.66\\\hline
         height (km) & $26.3\pm0.8$& $26.3\pm0.06$\\\hline
         velocity (km s$^{-1}$) & $3.5\pm0.3$& $3.8\pm0.1$\\\hline
         mass (kg)& $2.7 \pm 0.3$& $0.13\pm0.02$\\\hline
         shape density coefficient ($\kappa$ ; $ m^3kg^{-1}$) & $0.0032\pm0.0001$& $0.0039\pm0.0001$\\
        $\hookrightarrow$ density (kg m$^{-3}$); if $(A\times c_d) = 1.5$   & 3610 & 2650\\\hline
         ablation coefficient ($\sigma$ ; $s^2\,km^{-2}$)& $0.0141\pm0.00003$ & $0.0192\pm0.0003$\\\hline
    \end{tabular}
    \caption{Trajectory characteristics, including state values, estimated using the 3D particle filter for the final observation time of both events \textit{DN151212\_03} and \textit{DN160410\_03}.}
    \label{tab:3dpf_tf}
\end{table}

\subsubsection{Case 2: DN160410\_03}

\begin{figure}
    \centering
    \includegraphics[width=\textwidth]{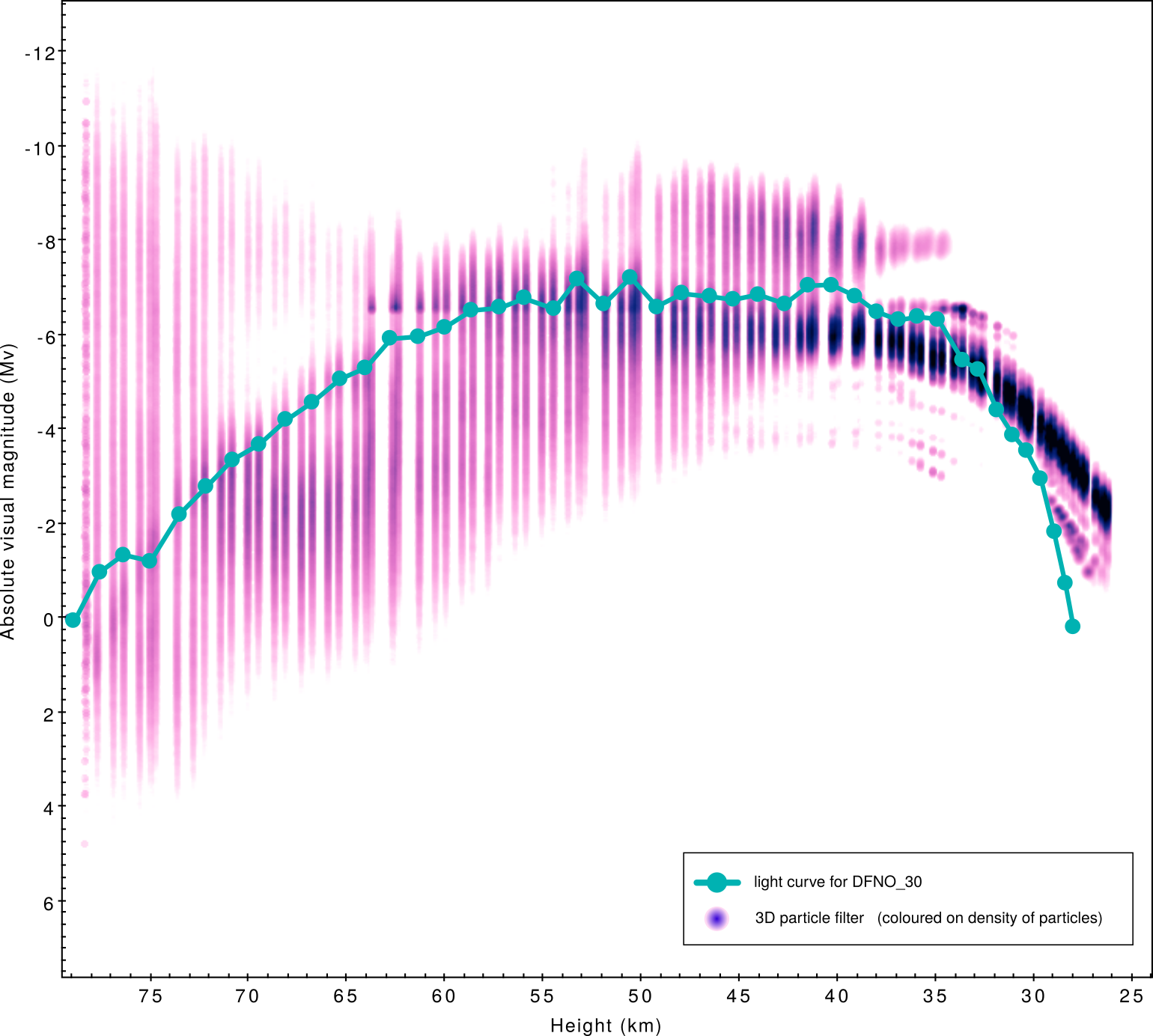}
    \caption{Comparison of light curve obtained from the DFNO\_30 still imagery (green) and predicted absolute visual magnitudes from 3D particle filter (coloured based on density of particles). The inferior match of higher weighted particles to the light curve toward the end can be attributed to the relatively constant value of $\tau$ (around $\sim0.2\%$) used for the calculation of predicted $M_v$ values (Equations \eqref{eqn:I}-\eqref{eqn:Iv}).}
    \label{fig:light}
\end{figure}

\begin{figure}
    \centering
    \includegraphics[width=\textwidth]{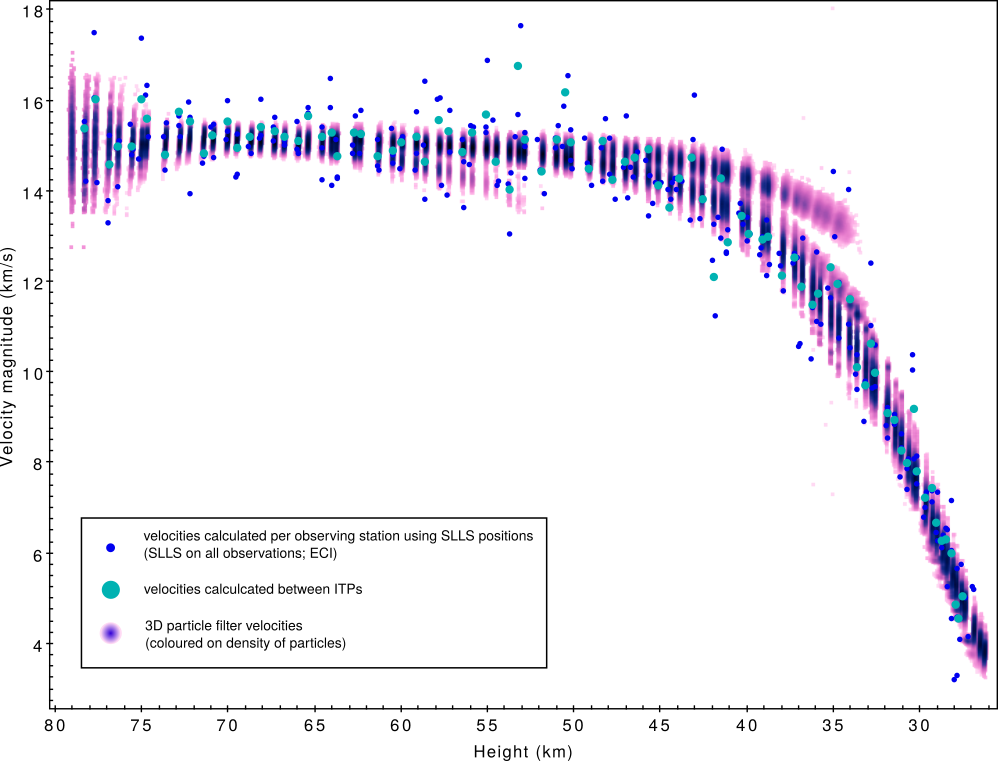}
    \caption{Magnitude of the velocity vector as calculated by the change in ITP positions with time (green), change in straight line least squares (SLLS) triangulated positions for each observatory with time (blue), and estimated by the 3D particle filter. }
    \label{fig:ba_vels}
\end{figure}

\begin{figure}
    \centering
    \includegraphics[width=\textwidth]{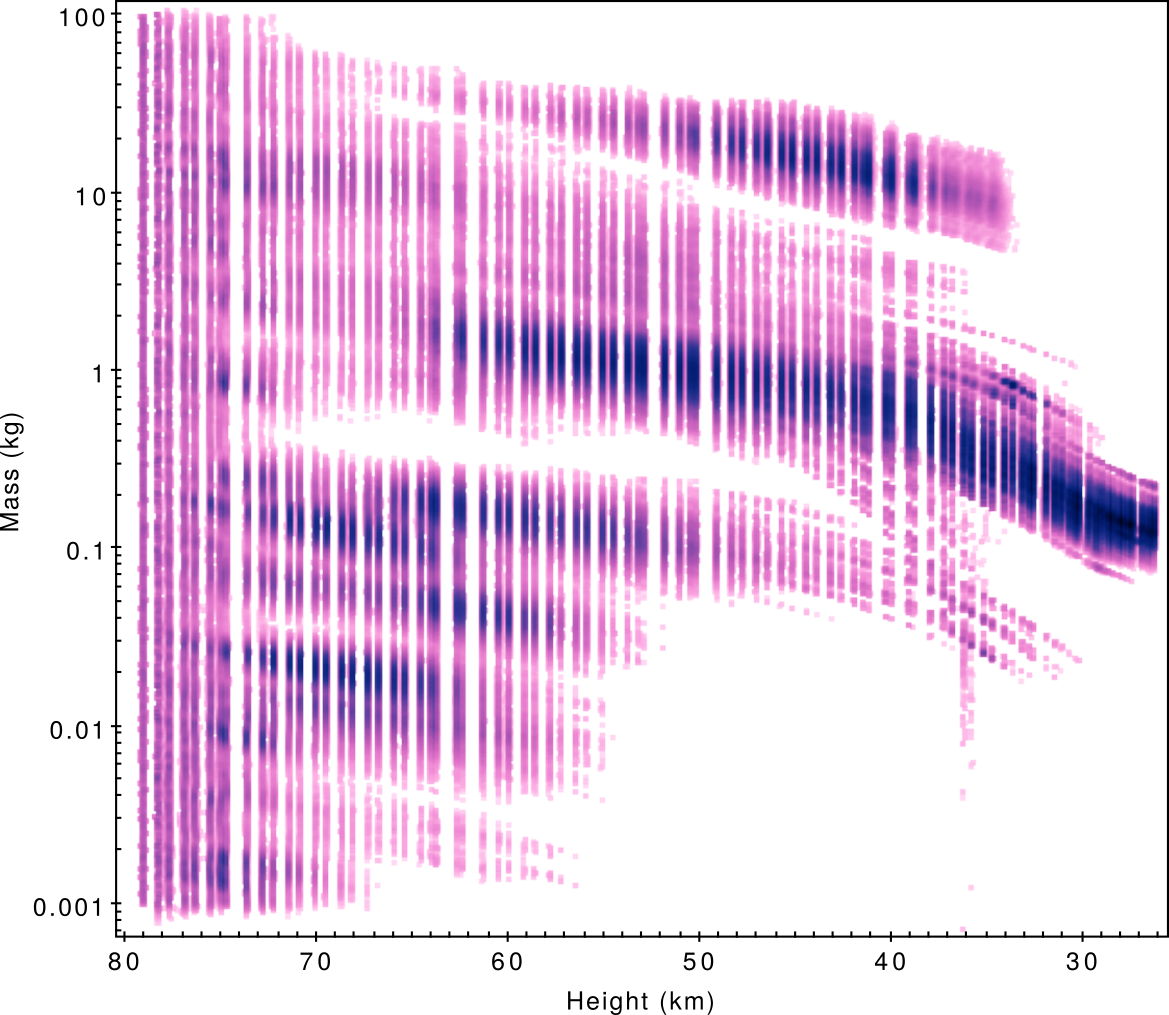}
    \caption{Masses estimated by the 3D particle filter, coloured by density of particles}
    \label{fig:ba_mass}
\end{figure}

The mean particle positions for event \textit{DN160410\_03} show a maximum deviation of $150\,m$, with nearly $80\%$ within $50\,m$ of the ITPs (Figure \ref{fig:ba_diffs}; black). 
Not only do the position estimates match the observations well, the calculated values of $M_v$ (evaluated using Equations \eqref{eqn:I} and \eqref{eqn:Iv}) also correspond well to the calibrated light curve for DFNO\_30 (Figure \ref{fig:light}). The inferior weightings attributed toward the end are most likely due to the calculation using the relatively constant value of $\tau$ (around $\sim0.2\%$). The good correlation between position and luminosity estimates to observational data validates the results of the particle filter, giving confidence to the estimates determined for  other state variables through the link in the state equations. The velocities for example can also be compared to those calculated between ITPs and the SLLS positions (Figure \ref{fig:ba_vels}). The exploration of this state space is interesting to observe. For example, we can see that a lower velocity option was tested at $\sim55$\,km but discontinues; a high velocity option around 35\,km experiences a similar fate. These discontinued streams can be linked to different mass options (Figure \ref{fig:ba_mass}).  The final meteoroid states for this exent at $t_f=4.6$ are also given in Table \ref{tab:3dpf_tf}.

\section{Discussion}

Fireball trajectories are typically approximated as straight line paths over a spherical Earth \citep{Ceplecha2005}.
This may be a reasonable assumption for short meteors, but for fireballs, effects that cause deviations to a straight line trajectory are not always negligible.
The astrometric uncertainty on DFN observations is typically $<1$ arcminute. This high precision, when propagated by the observational range to the fireball, gives uncertainties $\sim100$\,m. Any disturbances to the body greater than this will be resolvable.
Gravity and Earth rotation have known effects on trajectories and their observations respectively, and can be quantified. 
The $231$\,km long trajectory of the shallow event \textit{DN151212\_03} ($\gamma_e =17^{\circ}$) was observed for $21.14$\,s. This means a $>2.1$\,km downward displacement was experienced due to gravity alone. Over this length of time, at the latitude of the event, an observer on the ground would have moved nearly $8.5$\,km eastward with Earth's rotation. This must be accounted for if reduction is done in a non-inertial reference frame.
Event \textit{DN160410\_03} was steeper ($\gamma_e =65^{\circ}$) and significantly shorter in both duration (observed for 4.66 seconds) and length (58 km). Gravity therefore contributes a $105$\,m vertical displacement. The ground stations will also have moved 1.9\,km eastward, affecting apparent velocity vectors in a non-inertial frame. 

Fitting a linear trajectory to observations of a meteoroid will reduce the overall effects of gravity (and Earth rotation if using a non-inertial frame) by essentially averaging them out. This may provide usable position data, but will translate into a strong misrepresentation of velocity vectors. The difference in entry radiants calculated in both an inertial (ECI) and non-inertial (ECEF) frame for the upper trajectories demonstrates the effect of Earth rotation on these entry vectors. For event \textit{DN151212\_03} they vary by 1.8$^\circ$ and for event \textit{DN160410\_03} by 1.5$^\circ$.
Entry radiants are used in the calculation of fireball orbits. Integrating the fireball's motion back in time, performed to determine the heliocentric orbit beyond Earth's sphere of influence, is highly sensitive to both radiant direction and entry velocity. 
Using an inappropriate model to fit the observations introduces systematic errors to radiant angles, and as shown in Tables \ref{tab:bv_slls_tops} and \ref{tab:ba_slls_tops}, can be far greater than the quoted uncertainties based on the residual fit.
These systematic errors will affect orbit calculations, resulting in the incorrect evaluation  of a meteoroid's orbit.
The final velocity vector is used in dark flight modelling to estimate meteorite fall positions and will have similar issues, perhaps event more pronounced due to the lower velocities toward the end of the luminous trajectory.
Without a means of further testing meteoroid positions, there could be other forces involved that cause unmodelled deviations to a meteoroid trajectory.

Here we use individually triangulated positions (ITPs) as a base for comparison between meteoroid positions calculated using a straight line least squares approximation and a 3D particle filter. The calculation of the ITPs is a unique capability of the DFN as a result of absolute synchronisation of the time encoding between observatories. 
We have shown for both cases presented that there is a significant deviation of the meteoroid body when comparing ITPs to a straight line trajectory. 
To some degree the non-linear variability of these fireball trajectories can be visualised 
in Figures \ref{fig:bv_downline} and \ref{fig:ba_downline}. 
The absolute difference between the ITPs and the SLLS results are quantified for event \textit{DN151212\_03} in Figure \ref{fig:bv_diffs} and for event \textit{DN160410\_03}  in Figure \ref{fig:ba_diffs}. 
For both the long, shallow case (\textit{DN151212\_03}), and the steeper, shorter case (\textit{DN160410\_03}), the straight line trajectory does not represent the data well. For the triangulated positions using a straight line fitted to the entire data set, positions diverge up to $3.09\,km$ for the former and up to $360\,m$ for the latter.
The straight line trajectories fitted to data segmented at $50\, km$ give improved results for \textit{DN151212\_03} positions, diverging up to $750\,m$ for the upper trajectory, and $810\, m$ for the lower trajectory (discarding the $1.40\,km$ outlier at $49.2\,km$). The segmented triangulations for \textit{DN160410\_03} show an improvement only in the upper trajectory ($290\,m$), with an increased distance to the ITPs in the lower segment ($420\,m$). 
These deviations show that factors other than deceleration and ablation are able to significantly influence meteoroid trajectories. These could include aerodynamic effects on non-spherical bodies and, where fragmentation occurs, the dynamics involved in body disruption. 

We can approximate the magnitude of the forces required to cause the deviations seen in Figures \ref{fig:bv_downline} and \ref{fig:ba_downline}. Over the final 10 seconds of the \textit{DN151212\_03} trajectory (below 50 km), there is a lateral displacement of $\sim2.1$\,km. This results in an Eastward acceleration of $\sim42$\,m\,s$^{-2}$. For a 10\,kg body (minimum estimated mass at 50\,km altitude), this requires a lateral force of 420\,N. 
The vertical displacement of $\sim1.2$\,km seen in Figure \ref{fig:bv_downline} does not include any downward gravity component. As a vertical force would also have to overcome gravity to cause this change, an additional 460\,m displacement should be included (gravitational displacement normal to the trajectory over final 10 seconds =480\,m$\times \cos\gamma$). This gives a $\sim$330\,N upward force. 
Although \textit{DN160410\_03} wanders less drastically, a $\sim230$\,m lateral displacement from an altitude of 35\,km (the final 1.3\,seconds) requires a greater than 500\,N lateral force for a 2\,kg body (minimum estimated mass at 35\,km). From 30.2-27.3\,km (the final 0.7\,seconds), a vertical displacement normal to the trajectory of $\sim170$\,m would require over a 1000\,N  force. 

Work has been presented in the past on unique Earth-grazing events where a significant effort has been made to determine the path of the meteoroid without the unique use of a SLLS approximation \citep{borovicka1992graze, madiedo2016}. It is interesting to note however that in \citet{borovicka1992graze} there is an observatory almost directly under the event from which the authors were able to determine that there was no curvature to the trajectory outside the observational plane from this viewpoint. For event \textit{DN151212\_03} analysed here, there was a deviation from the SLLS trajectory not only in altitude, but with a significant lateral component. 
Because of its large size and extreme ablation duration, \textit{DN151212\_03}  may not be a typical event, however \textit{DN160410\_03} is an ideal example of a meteorite dropping fireball. The deviation of the \textit{DN160410\_03} fireball from a straight line shows that an SLLS may not be an appropriate approximation for the majority of deep-penetrating ($<50$\,km altitude) fireballs. 
The cross-track forces as approximated above, are certainly significant, complicating the ideal straight line scenario and bringing into question the reliability of using this assumption even for small events. Their origins, be they aerodynamic, related to fragmentation or as yet unconsidered, should be investigated.

The complexity of meteoroid trajectories makes it difficult to simulate them with simplified model equations such as given by Equation \eqref{eqn:components}. 
Using this single dimension model in a particle filter (e.g. \citet{Sansom2017}) forces the measurement update step to use straight line position values for distance travelled along the trajectory. 
This misrepresentation of the data in the filter can not only affect position estimates, but may additionally influence other state parameters through the relationship in the state equations  \eqref{eqn:components}, such as velocity and mass values. 
As the particle filter is an adaptive approach that uses observations to update state estimates, using the most unprocessed measurements permits subtleties in the data to influence the predicted state.
Using the three dimensional model \eqref{eqn:components}, it is able to use the raw line-of-sight observations as described in Section \ref{sec:meas}. 
Using a 3D particle filter also provides a more robust error analysis as uncertainties are propagated comprehensively from well constrained astrometric errors through to the end of the luminous trajectory. 
At $t_{end}$, the remaining particles can be used as a direct input to Monte Carlo dark flight simulations, as presented by \citet{2018arXiv180302557D}.
The minimisation of time spent in the field searching for meteorites is of great importance. 
It is therefore essential to define a search region on the ground that is representative of the statistical results obtained from physical modelling of bright flight observations.
The final mass given using an extended Kalman smoother on pre-triangulated straight line data for event \textit{DN151212\_03} is $2.0\pm0.2$\,kg (Table \ref{tab:bv_slls}) and $0.05\pm0.01$ (Table \ref{tab:ba_slls}) for event \textit{DN160410\_03}, compared to the 2.7$\pm$0.3 kg and 0.13$\pm$0.02 kg final masses predicted for these events using the particle filter. 
Identifying events with greater chances of a successful find will significantly influence decisions about the feasibility of a remote search for a given event.
Shallow events in particular, such as \textit{DN151212\_03} ($\gamma_e =15.8^{\circ}$) tend to produce extended fall lines, tens of kilometres long, from small fragments to main body masses. Well constrained final states in these cases are essential. 

\section{Conclusion}
As fireball producing events are typically associated with larger asteroidal debris they have the ability to penetrate deep into the Earth's atmosphere. These events can last tens of seconds, with ground based observations influenced by Earth's rotation and gravity effects resolvable with modern camera resolution. The unique ability of the Desert Fireball Network to triangulate a meteoroid's position at discrete times allows us to investigate the true variability of trajectories. These individually triangulated positions (ITPs) are used as a reference for comparison to other methods of evaluating meteoroid positions. 
The flights of two fireballs observed by the Desert Fireball Network were investigated as example events.
Triangulating data using a straight line assumption eliminates subtleties in the data that may be indicative of unmodelled processes, such as fragmentation and aerodynamic effects. 
Deviations from a straight line path of up to $3.09\,km$ for event \textit{DN151212\_03} and $360\,m$ for event \textit{DN160410\_03} were observed, and a downline view in an inertial reference frame (ECI) shows this is mostly lateral. The investigation in an ECI reference frame eliminates Earth rotation effects, and, as these deviations cannot be accounted for by gravity, must have a different cause. Even the more typical event \textit{DN160410\_03} is affected, showing all influences on fireball trajectories should be considered in all deep penetrating cases. The misrepresentation of the start and end of meteoroid trajectories by a straight line fit will affect dark flight models for meteorite search regions as well as orbit determination.

Modelling fireball camera network data in three dimensions has not previously been attempted. 
The self-contained particle filter approach of \citet{Sansom2017} has been adapted to use a three dimensional dynamic model, and incorporate absolute visual magnitude observations. This allows the raw astrometric observations as seen by each observatory to be incorporated directly into the estimation of a meteoroid state, removing the need for pre-triangulated measurement data.
By incorporating the raw observations, errors in each azimuth and elevation can be accounted for and propagated individually. This results in a final state estimate with fully comprehensive errors, leading to more realistic meteorite search areas and will allow an automated, systematic evaluation of trajectories observed by multiple station camera networks.

\section{Acknowledgements}
This work was funded by the Australian Research Council as part of the Australian Discovery Project scheme, and supported by resources provided by the Pawsey Supercomputing Centre with funding from the Australian Government and the Government of Western Australia. This research made use of Astropy, a community-developed core Python package for Astronomy \citep{robitaille2013astropy} and figures were generated for the large resulting dataset using TOPCAT \citep{taylor2005topcat}.

    \bibliography{Bibliography}{}

\begin{thebibliography}{34}
\providecommand{\natexlab}[1]{#1}
\providecommand{\url}[1]{\texttt{#1}}
\expandafter\ifx\csname urlstyle\endcsname\relax
  \providecommand{\doi}[1]{doi: #1}\else
  \providecommand{\doi}{doi: \begingroup \urlstyle{rm}\Url}\fi

\bibitem[Artemieva and Pierazzo(2009)]{Artemieva2009}
N.~Artemieva and E.~Pierazzo.
\newblock {The Canyon Diablo impact event: Projectile motion through the
  atmosphere}.
\newblock \emph{Meteoritics {\&} Planetary Science}, 44\penalty0 (1):\penalty0
  25--42, 2009.
\newblock ISSN 1945-5100.
\newblock \doi{10.1111/j.1945-5100.2009.tb00715.x}.
\newblock URL \url{http://dx.doi.org/10.1111/j.1945-5100.2009.tb00715.x}.

\bibitem[{Artemieva} and {Shuvalov}(2016)]{Artemieva2016}
N.~A. {Artemieva} and V.~V. {Shuvalov}.
\newblock {From Tunguska to Chelyabinsk via Jupiter}.
\newblock \emph{Annual Review of Earth and Planetary Sciences}, 44:\penalty0
  37--56, 2016.
\newblock \doi{10.1146/annurev-earth-060115-012218}.

\bibitem[Baldwin and Sheaffer(1971)]{Baldwin1971}
B.~Baldwin and Y.~Sheaffer.
\newblock Ablation and breakup of large meteoroids during atmospheric entry.
\newblock \emph{Journal of Geophysical Research}, 76\penalty0 (19):\penalty0
  4653--4668, 1971.

\bibitem[Borovi{\v{c}}ka(1990)]{Borovicka1990}
J.~Borovi{\v{c}}ka.
\newblock The comparison of two methods of determining meteor trajectories from
  photographs.
\newblock \emph{Bulletin of the Astronomical Institutes of Czechoslovakia},
  41:\penalty0 391--396, 1990.

\bibitem[{Borovicka} and {Ceplecha}(1992)]{borovicka1992graze}
J.~{Borovicka} and Z.~{Ceplecha}.
\newblock {Earth-grazing fireball of October 13, 1990}.
\newblock \emph{AAP}, 257:\penalty0 323--328, Apr. 1992.

\bibitem[Borovi{\v{c}}ka et~al.(2013)Borovi{\v{c}}ka, T{\'o}th, Igaz,
  Spurn{\'y}, Kalenda, Haloda, Svore{\v{n}}, Korno{\v{s}}, Silber, Brown,
  et~al.]{Borovicka2013kosice}
J.~Borovi{\v{c}}ka, J.~T{\'o}th, A.~Igaz, P.~Spurn{\'y}, P.~Kalenda, J.~Haloda,
  J.~Svore{\v{n}}, L.~Korno{\v{s}}, E.~Silber, P.~Brown, et~al.
\newblock {The Ko{\v{s}}ice meteorite fall: Atmospheric trajectory,
  fragmentation, and orbit}.
\newblock \emph{Meteoritics \& Planetary Science}, 48\penalty0 (10):\penalty0
  1757--1779, 2013.
\newblock ISSN 10869379.

\bibitem[Borovi{\v{c}}ka et~al.(2015)Borovi{\v{c}}ka, Spurn{\'{y}},
  {\v{S}}egon, Andrei{\'{c}}, Kac, Korlevi{\'{c}}, Atanackov, Kladnik, Mucke,
  Vida, and Novoselnik]{Borovicka2015Krizevci}
J.~Borovi{\v{c}}ka, P.~Spurn{\'{y}}, D.~{\v{S}}egon, {\v{Z}}.~Andrei{\'{c}},
  J.~Kac, K.~Korlevi{\'{c}}, J.~Atanackov, G.~Kladnik, H.~Mucke, D.~Vida, and
  F.~Novoselnik.
\newblock {The instrumentally recorded fall of the Kri{\{}{\v{z}}{\}}evci
  meteorite, Croatia, February 4, 2011}.
\newblock \emph{Meteoritics {\&} Planetary Science}, 16\penalty0 (7), 2015.
\newblock \doi{10.1111/maps.12469}.

\bibitem[Bronshten(1983)]{Bronshten1983}
V.~A. Bronshten.
\newblock \emph{{Physics of Meteoric Phenomena}}.
\newblock Geophysics and Astrophysics Monographs. Reidel, Dordrecht,
  Netherlands, 1983.
\newblock ISBN 9789027716545.

\bibitem[Brown et~al.(2011)Brown, McCausland, Fries, Silber, Edwards, Wong,
  Weryk, Fries, and Krzeminski]{Brown2011}
P.~Brown, P.~J.~A. McCausland, M.~Fries, E.~Silber, W.~N. Edwards, D.~K. Wong,
  R.~J. Weryk, J.~Fries, and Z.~Krzeminski.
\newblock The fall of the grimsby meteorite—i: Fireball dynamics and orbit
  from radar, video, and infrasound records.
\newblock \emph{Meteoritics \& Planetary Science}, 46\penalty0 (3):\penalty0
  339--363, 2011.
\newblock ISSN 1945-5100.
\newblock \doi{10.1111/j.1945-5100.2010.01167.x}.
\newblock URL \url{http:https://dx.doi.org/10.1111/j.1945-5100.2010.01167.x}.

\bibitem[Brown et~al.(1994)Brown, Ceplecha, Hawkes, Wetherill, Beech, and
  Mossman]{Brown1994}
P.~G. Brown, Z.~Ceplecha, R.~L. Hawkes, G.~Wetherill, M.~Beech, and K.~Mossman.
\newblock {The orbit and atmospheric trajectory of the Peekskill meteorite from
  video records}.
\newblock \emph{Nature}, 367\penalty0 (6464):\penalty0 624--626, 1994.
\newblock ISSN 0028-0836.
\newblock \doi{10.1038/367624a0}.

\bibitem[Campbell-Brown and Koschny(2004)]{Campbell-Brown2004}
M.~D. Campbell-Brown and D.~Koschny.
\newblock {Model of the ablation of faint meteors}.
\newblock \emph{Astronomy and Astrophysics}, 418\penalty0 (2):\penalty0
  751--758, 2004.
\newblock ISSN 0004-6361.
\newblock \doi{10.1051/0004-6361:20041001-1}.

\bibitem[Ceplecha(1987)]{Ceplecha1987}
Z.~Ceplecha.
\newblock Geometric, dynamic, orbital and photometric data on meteoroids from
  photographic fireball networks.
\newblock \emph{Bulletin of the Astronomical Institutes of Czechoslovakia},
  38:\penalty0 222--234, 1987.

\bibitem[Ceplecha and Revelle(2005)]{Ceplecha2005}
Z.~Ceplecha and D.~O. Revelle.
\newblock {Fragmentation model of meteoroid motion, mass loss, and radiation in
  the atmosphere}.
\newblock \emph{Meteoritics {\&} Planetary Science}, 40\penalty0 (1):\penalty0
  35--54, 2005.
\newblock ISSN 10869379.
\newblock \doi{10.1111/j.1945-5100.2005.tb00363.x}.

\bibitem[Ceplecha et~al.(1996)Ceplecha, Spalding, Jacobs, and
  Tagliaferri]{Ceplecha1996}
Z.~Ceplecha, R.~E. Spalding, C.~F. Jacobs, and E.~Tagliaferri.
\newblock Luminous efficiencies of bolides.
\newblock In \emph{SPIE's 1996 International Symposium on Optical Science,
  Engineering, and Instrumentation}, volume 2813, pages 46--56. International
  Society for Optics and Photonics, 1996.
\newblock \doi{10.1117/12.256072}.

\bibitem[Ceplecha et~al.(1998)Ceplecha, Borovi{\v{c}}ka, Elford, ReVelle,
  Hawkes, Porub{\v{c}}an, and {\v{S}}imek]{Ceplecha1998}
Z.~Ceplecha, J.~Borovi{\v{c}}ka, W.~G. Elford, D.~O. ReVelle, R.~L. Hawkes,
  V.~Porub{\v{c}}an, and M.~{\v{S}}imek.
\newblock {Meteor Phenomena and Bodies}.
\newblock \emph{Space Science Reviews}, 84\penalty0 (3):\penalty0 327--471,
  1998.
\newblock ISSN 1572-9672.
\newblock \doi{10.1023/A:1005069928850}.

\bibitem[{Devillepoix} et~al.(2018){Devillepoix}, {Sansom}, {Bland}, {Towner},
  {Cup{\'a}k}, {Howie}, {Jansen-Sturgeon}, {Cox}, {Hartig}, {Benedix}, and
  {Paxman}]{2018arXiv180302557D}
H.~A.~R. {Devillepoix}, E.~K. {Sansom}, P.~A. {Bland}, M.~C. {Towner},
  M.~{Cup{\'a}k}, R.~M. {Howie}, T.~{Jansen-Sturgeon}, M.~A. {Cox}, B.~A.~D.
  {Hartig}, G.~K. {Benedix}, and J.~P. {Paxman}.
\newblock {The Dingle Dell meteorite: a Halloween treat from the Main Belt}.
\newblock \emph{ArXiv e-prints}, Mar. 2018.

\bibitem[Grewal and Andrews(1993)]{Grewal1993}
M.~S. Grewal and A.~P. Andrews.
\newblock \emph{{Kalman filtering: theory and practice}}.
\newblock Prentice-Hall Inc., New Jersey, 1993.

\bibitem[Hildebrand et~al.(2006)Hildebrand, McCausland, Brown, Longstaffe,
  Russell, Tagliaferri, Wacker, and Mazur]{Hildebrand2006}
A.~R. Hildebrand, P.~J.~A. McCausland, P.~G. Brown, F.~J. Longstaffe, S.~D.~J.
  Russell, E.~Tagliaferri, J.~F. Wacker, and M.~J. Mazur.
\newblock The fall and recovery of the tagish lake meteorite.
\newblock \emph{Meteoritics {\&} Planetary Science}, 41\penalty0 (3):\penalty0
  407--431, 2006.
\newblock ISSN 1945-5100.
\newblock \doi{10.1111/j.1945-5100.2006.tb00471.x}.
\newblock URL \url{http://dx.doi.org/10.1111/j.1945-5100.2006.tb00471.x}.

\bibitem[Hoppe(1937)]{Hoppe1937}
J.~Hoppe.
\newblock {Die physikalischen Vorg{\"{a}}nge beim Eindringen meteoritischer
  K{\"{o}}rper in die Erdatmosph{\"{a}}re}.
\newblock \emph{Astronomische Nachrichten}, 262\penalty0 (10):\penalty0
  169--198, 1937.

\bibitem[Howie et~al.(2017)Howie, Paxman, Bland, Towner, Sansom, and
  Devillepoix]{howie2017deb}
R.~M. Howie, J.~Paxman, P.~A. Bland, M.~C. Towner, E.~K. Sansom, and H.~A.
  Devillepoix.
\newblock Submillisecond fireball timing using de bruijn timecodes.
\newblock \emph{Meteoritics \& Planetary Science}, 2017.

\bibitem[Kikwaya et~al.(2011)Kikwaya, Campbell-Brown, and Brown]{Kikwaya2011}
J.-B. Kikwaya, M.~D. Campbell-Brown, and P.~G. Brown.
\newblock {Bulk density of small meteoroids}.
\newblock \emph{Astronomy {\&} Astrophysics}, 530:\penalty0 A113, 2011.
\newblock ISSN 0004-6361.
\newblock \doi{10.1051/0004-6361/201116431}.

\bibitem[Madiedo et~al.(2016)Madiedo, Espartero, Castro-Tirado, Pastor, and
  de~los Reyes]{madiedo2016}
J.~M. Madiedo, F.~Espartero, A.~J. Castro-Tirado, S.~Pastor, and J.~A. de~los
  Reyes.
\newblock An earth-grazing fireball from the daytime ζ-perseid shower observed
  over spain on 2012 june 10.
\newblock \emph{Monthly Notices of the Royal Astronomical Society},
  460\penalty0 (1):\penalty0 917--922, 2016.
\newblock \doi{10.1093/mnras/stw1020}.
\newblock URL \url{+ http://dx.doi.org/10.1093/mnras/stw1020}.

\bibitem[McCrosky and Boeschenstein(1965)]{McCrosky1965}
R.~E. McCrosky and H.~Boeschenstein.
\newblock {The Prairie Meteorite Network}.
\newblock \emph{Optical Engineering}, 3\penalty0 (4):\penalty0 304127--304127,
  1965.

\bibitem[Picone et~al.(2002)Picone, Hedin, Drob, and Aikin]{Picone2002}
J.~M. Picone, A.~E. Hedin, D.~P. Drob, and A.~C. Aikin.
\newblock {NRLMSISE-00 empirical model of the atmosphere: Statistical
  comparisons and scientific issues}.
\newblock \emph{Journal of Geophysical Research: Space Physics (1978--2012)},
  107\penalty0 (A12):\penalty0 1468, 2002.

\bibitem[Ristic et~al.(2004)Ristic, Arulampalam, and Gordon]{Ristic2004}
B.~Ristic, S.~Arulampalam, and N.~Gordon.
\newblock \emph{Beyond the Kalman filter: Particle filters for tracking
  applications}, volume 685.
\newblock Artech house Boston, 2004.

\bibitem[Robitaille et~al.(2013)Robitaille, Tollerud, Greenfield, Droettboom,
  Bray, Aldcroft, Davis, Ginsburg, Price-Whelan, Kerzendorf,
  et~al.]{robitaille2013astropy}
T.~P. Robitaille, E.~J. Tollerud, P.~Greenfield, M.~Droettboom, E.~Bray,
  T.~Aldcroft, M.~Davis, A.~Ginsburg, A.~M. Price-Whelan, W.~E. Kerzendorf,
  et~al.
\newblock Astropy: A community python package for astronomy.
\newblock \emph{Astronomy \& Astrophysics}, 558:\penalty0 A33, 2013.

\bibitem[Sansom et~al.(2017)Sansom, Rutten, and Bland]{Sansom2017}
E.~Sansom, M.~Rutten, and P.~Bland.
\newblock Analyzing meteoroid flights using particle filters.
\newblock \emph{The Astronomical Journal}, 153\penalty0 (2):\penalty0 87, 2017.

\bibitem[Sansom et~al.(2015)Sansom, Bland, Paxman, and Towner]{Sansom2015}
E.~K. Sansom, P.~A. Bland, J.~Paxman, and M.~C. Towner.
\newblock {A novel approach to fireball modeling: The observable and the
  calculated}.
\newblock \emph{Meteoritics {\&} Planetary Science}, 50\penalty0 (8):\penalty0
  1423--1435, 2015.
\newblock ISSN 10869379.
\newblock \doi{10.1111/maps.12478}.

\bibitem[Shuvalov(1999)]{Shuvalov1999}
V.~Shuvalov.
\newblock Multi-dimensional hydrodynamic code sova for interfacial flows:
  Application to the thermal layer effect.
\newblock \emph{Shock Waves}, 9\penalty0 (6):\penalty0 381--390, 1999.
\newblock ISSN 1432-2153.
\newblock \doi{10.1007/s001930050168}.
\newblock URL \url{http://dx.doi.org/10.1007/s001930050168}.

\bibitem[Shuvalov and Artemieva(2002)]{Shuvalov2002}
V.~V. Shuvalov and N.~a. Artemieva.
\newblock {Numerical modeling of Tunguska-like impacts}.
\newblock \emph{Planetary and Space Science}, 50\penalty0 (2):\penalty0
  181--192, 2002.
\newblock ISSN 00320633.
\newblock \doi{10.1016/S0032-0633(01)00079-4}.

\bibitem[Spurn{\'{y}} et~al.(2010)Spurn{\'{y}}, Borovi{\v{c}}ka, Kac, Kalenda,
  Atanackov, Kladnik, Heinlein, and Grau]{Spurny2010jen}
P.~Spurn{\'{y}}, J.~Borovi{\v{c}}ka, J.~Kac, P.~Kalenda, J.~Atanackov,
  G.~Kladnik, D.~Heinlein, and T.~Grau.
\newblock Analysis of instrumental observations of the jesenice meteorite fall
  on april 9, 2009.
\newblock \emph{Meteoritics \& Planetary Science}, 45\penalty0 (8):\penalty0
  1392--1407, 2010.
\newblock ISSN 1945-5100.
\newblock \doi{10.1111/j.1945-5100.2010.01121.x}.
\newblock URL \url{http:https://dx.doi.org/10.1111/j.1945-5100.2010.01121.x}.

\bibitem[Spurn{\'{y}} et~al.(2012)Spurn{\'{y}}, Bland, Shrben{\'{y}},
  Borovi{\v{c}}ka, Ceplecha, Singelton, Bevan, Vaughan, Towner, Mcclafferty,
  Toumi, and Deacon]{Spurny2012}
P.~Spurn{\'{y}}, P.~Bland, L.~Shrben{\'{y}}, J.~Borovi{\v{c}}ka, Z.~Ceplecha,
  A.~Singelton, A.~W.~R. Bevan, D.~Vaughan, M.~C. Towner, T.~P. Mcclafferty,
  R.~Toumi, and G.~Deacon.
\newblock {The Bunburra Rockhole meteorite fall in SW Australia: Fireball
  trajectory, luminosity, dynamics, orbit, and impact position from
  photographic and photoelectric records}.
\newblock \emph{Meteoritics {\&} Planetary Science}, 47\penalty0 (2):\penalty0
  163--185, 2012.
\newblock \doi{10.1111/j.1945-5100.2011.01321.x}.

\bibitem[Spurn{\'{y}} et~al.(2017)Spurn{\'{y}}, Borovi{\v{c}}ka, Baumgarten,
  Haack, Heinlein, and S{\o}rensen]{Spurny2017Atmospheric2016}
P.~Spurn{\'{y}}, J.~Borovi{\v{c}}ka, G.~Baumgarten, H.~Haack, D.~Heinlein, and
  A.~N. S{\o}rensen.
\newblock {Atmospheric trajectory and heliocentric orbit of the Ejby meteorite
  fall in Denmark on February 6, 2016}.
\newblock \emph{Planetary and Space Science}, 2017.
\newblock ISSN 00320633.
\newblock \doi{10.1016/j.pss.2016.11.010}.

\bibitem[Taylor(2005)]{taylor2005topcat}
M.~B. Taylor.
\newblock {TOPCAT \& STILTS: starlink table/VOTable processing software}.
\newblock In \emph{Astronomical Data Analysis Software and Systems XIV}, volume
  347, page~29, 2005.

\end{thebibliography}
    \bibliographystyle{abbrvnat}

\end{document}